\input harvmac
\input epsf
\input amssym


\newcount\figno
\figno=0
\def\fig#1#2#3{
\par\begingroup\parindent=0pt\leftskip=1cm\rightskip=1cm\parindent=0pt
\baselineskip=11pt \global\advance\figno by 1 \midinsert
\epsfxsize=#3 \centerline{\epsfbox{#2}} \vskip 12pt {\bf Fig.\
\the\figno: } #1\par
\endinsert\endgroup\par
}
\def\figlabel#1{\xdef#1{\the\figno}}
\def\encadremath#1{\vbox{\hrule\hbox{\vrule\kern8pt\vbox{\kern8pt
\hbox{$\displaystyle #1$}\kern8pt} \kern8pt\vrule}\hrule}}

\def\ev#1{\langle#1\rangle}

\def\frac#1#2{{#1 \over #2}}

\def\p{\partial}
\def\semi{\subset\kern-1em\times\;}
\def\bar#1{\overline{#1}}
\def\sqr#1#2{{\vcenter{\vbox{\hrule height.#2pt
\hbox{\vrule width.#2pt height#1pt \kern#1pt \vrule width.#2pt}
\hrule height.#2pt}}}}

\def\IZ{\Bbb{Z}}
\def\IC{\Bbb{C}}

\def\IR{\Bbb{R}}

\def\Tr{ {\rm Tr}}
\def\p{\partial}
\def\go{g^{(0)}}
\def\wb{\overline{w}}
\def\Lt{\tilde{L}}
\def\ct{\tilde{c}}
\def\taub{\overline{\tau}}
\def\zb{\overline{z}}
\def\kt{\tilde{k}}
\def\A0{A^{(0)}}
\def\At{\tilde{A}}
\def\kt{\tilde{k}}
\def\Jt{\tilde{J}}
\def\qt{\tilde{q}}
\def\ap{\alpha'}
\def\zt{\tilde{z}}
\def\qb{\overline{q}}
\def\chit{\tilde{\chi}}
\def\etat{\tilde{\eta}}
\def\mut{\tilde{\mu}}


\lref\farey{
  R.~Dijkgraaf, J.~M.~Maldacena, G.~W.~Moore and E.~P.~Verlinde,
   ``A black hole farey tail,''
  arXiv:hep-th/0005003.
}

\lref\KrausNB{
  P.~Kraus and F.~Larsen,
  ``Partition functions and elliptic genera from supergravity,''
  arXiv:hep-th/0607138.
}

\lref\HansenWU{
  J.~Hansen and P.~Kraus,
  ``Generating charge from diffeomorphisms,''
  arXiv:hep-th/0606230.
}

\lref\KrausZM{
  P.~Kraus and F.~Larsen,
  ``Holographic gravitational anomalies,''
  JHEP {\bf 0601}, 022 (2006)
  [arXiv:hep-th/0508218].
}

\lref\KrausVZ{
  P.~Kraus and F.~Larsen,
  ``Microscopic black hole entropy in theories with higher derivatives,''
  JHEP {\bf 0509}, 034 (2005)
  [arXiv:hep-th/0506176].
}
\lref\KrausDI{
  P.~Kraus, F.~Larsen and R.~Siebelink,
  ``The gravitational action in asymptotically AdS and flat spacetimes,''
  Nucl.\ Phys.\ B {\bf 563}, 259 (1999)
  [arXiv:hep-th/9906127].
}

\lref\BalasubramanianRE{
  V.~Balasubramanian and P.~Kraus,
  ``A stress tensor for anti-de Sitter gravity,''
  Commun.\ Math.\ Phys.\  {\bf 208}, 413 (1999)
  [arXiv:hep-th/9902121].
}

\lref\deBoerVG{
  J.~de Boer, M.~C.~N.~Cheng, R.~Dijkgraaf, J.~Manschot and E.~Verlinde,
  ``A farey tail for attractor black holes,''
  arXiv:hep-th/0608059.
}

\lref\deBoerUS{
  J.~de Boer,
  ``Large N Elliptic Genus and AdS/CFT Correspondence,''
  JHEP {\bf 9905}, 017 (1999)
  [arXiv:hep-th/9812240].
}

\lref\deBoerIP{
  J.~de Boer,
  ``Six-dimensional supergravity on S**3 x AdS(3) and 2d conformal field
  theory,''
  Nucl.\ Phys.\ B {\bf 548}, 139 (1999)
  [arXiv:hep-th/9806104].
}

\lref\BeasleyUS{
  C.~Beasley, D.~Gaiotto, M.~Guica, L.~Huang, A.~Strominger and X.~Yin,
  ``Why Z(BH) = |Z(top)|**2,''
  arXiv:hep-th/0608021.
}

\lref\GaiottoNS{
  D.~Gaiotto, A.~Strominger and X.~Yin,
  ``From AdS(3)/CFT(2) to black holes / topological strings,''
  arXiv:hep-th/0602046.
}
\lref\OoguriZV{
  H.~Ooguri, A.~Strominger and C.~Vafa,
  ``Black hole attractors and the topological string,''
  Phys.\ Rev.\ D {\bf 70}, 106007 (2004)
  [arXiv:hep-th/0405146].
}

\lref\GukovID{
  S.~Gukov, E.~Martinec, G.~W.~Moore and A.~Strominger,
  ``Chern-Simons gauge theory and the AdS(3)/CFT(2) correspondence,''
  arXiv:hep-th/0403225.
}

\lref\MaldacenaBW{
  J.~M.~Maldacena and A.~Strominger,
  ``AdS(3) black holes and a stringy exclusion principle,''
  JHEP {\bf 9812}, 005 (1998)
  [arXiv:hep-th/9804085].
}

\lref\StromingerEQ{
  A.~Strominger,
  ``Black hole entropy from near-horizon microstates,''
  JHEP {\bf 9802}, 009 (1998)
  [arXiv:hep-th/9712251].
}
\lref\MaldacenaDE{
  J.~M.~Maldacena, A.~Strominger and E.~Witten,
  JHEP {\bf 9712}, 002 (1997)
  [arXiv:hep-th/9711053].
}

\lref\StromingerSH{
  A.~Strominger and C.~Vafa,
  ``Microscopic Origin of the Bekenstein-Hawking Entropy,''
  Phys.\ Lett.\ B {\bf 379}, 99 (1996)
  [arXiv:hep-th/9601029].
}

\lref\finn{
  F.~Larsen,
  ``The perturbation spectrum of black holes in N = 8 supergravity,''
  Nucl.\ Phys.\ B {\bf 536}, 258 (1998)
  [arXiv:hep-th/9805208].
  }

\lref\GopakumarII{
  R.~Gopakumar and C.~Vafa,
  ``M-theory and topological strings. I,''
  arXiv:hep-th/9809187.  ``M-theory and topological strings. II,''
  arXiv:hep-th/9812127.
}

\lref\MathurZP{
  S.~D.~Mathur,
  ``The fuzzball proposal for black holes: An elementary review,''
  Fortsch.\ Phys.\  {\bf 53}, 793 (2005)
  [arXiv:hep-th/0502050].
}

\lref\BalasubramanianRT{
  V.~Balasubramanian, J.~de Boer, E.~Keski-Vakkuri and S.~F.~Ross,
  ``Supersymmetric conical defects: Towards a string theoretic description  of
  black hole formation,''
  Phys.\ Rev.\ D {\bf 64}, 064011 (2001)
  [arXiv:hep-th/0011217].
}

\lref\MaldacenaDR{
  J.~M.~Maldacena and L.~Maoz,
  ``De-singularization by rotation,''
  JHEP {\bf 0212}, 055 (2002)
  [arXiv:hep-th/0012025];
  O.~Lunin, J.~M.~Maldacena and L.~Maoz,
  ``Gravity solutions for the D1-D5 system with angular momentum,''
  arXiv:hep-th/0212210;
  }

\lref\KutasovZH{
  D.~Kutasov, F.~Larsen and R.~G.~Leigh,
  ``String theory in magnetic monopole backgrounds,''
  Nucl.\ Phys.\ B {\bf 550}, 183 (1999)
  [arXiv:hep-th/9812027].
}

\lref\generalRRRR{
  M.~B.~Green and J.~H.~Schwarz,
  ``Supersymmetrical Dual String Theory. 2. Vertices And Trees,''
  Nucl.\ Phys.\ B {\bf 198}, 252 (1982);
  D.~J.~Gross and E.~Witten,
  ``Superstring Modifications Of Einstein's Equations,''
  Nucl.\ Phys.\ B {\bf 277}, 1 (1986);
  W.~Lerche, B.~E.~W.~Nilsson and A.~N.~Schellekens,
  ``Heterotic String Loop Calculation Of The Anomaly Cancelling Term,''
  Nucl.\ Phys.\ B {\bf 289}, 609 (1987);
  M.~J.~Duff, J.~T.~Liu and R.~Minasian,
  ``Eleven-dimensional origin of string / string duality: A one-loop test,''
  Nucl.\ Phys.\ B {\bf 452}, 261 (1995)
  [arXiv:hep-th/9506126];
   M.~B.~Green, M.~Gutperle and P.~Vanhove,
  ``One loop in eleven dimensions,''
  Phys.\ Lett.\ B {\bf 409}, 177 (1997)
  [arXiv:hep-th/9706175];
    J.~G.~Russo and A.~A.~Tseytlin,
  ``One-loop four-graviton amplitude in eleven-dimensional supergravity,''
  Nucl.\ Phys.\ B {\bf 508}, 245 (1997)
  [arXiv:hep-th/9707134]; P.~S.~Howe and D.~Tsimpis,
  ``On higher-order corrections in M theory,''
  JHEP {\bf 0309}, 038 (2003)
  [arXiv:hep-th/0305129].
}

\lref\KrausGH{
  P.~Kraus and F.~Larsen,
  ``Attractors and black rings,''
  Phys.\ Rev.\ D {\bf 72}, 024010 (2005)
  [arXiv:hep-th/0503219].
}

\lref\HMM{ D.~Freed, J.~A.~Harvey, R.~Minasian and G.~W.~Moore,
  ``Gravitational anomaly cancellation for M-theory fivebranes,''
  Adv.\ Theor.\ Math.\ Phys.\  {\bf 2}, 601 (1998)
  [arXiv:hep-th/9803205];
J.~A.~Harvey, R.~Minasian and G.~W.~Moore,
``Non-abelian tensor-multiplet anomalies,''
 JHEP {\bf 9809}, 004 (1998)
  [arXiv:hep-th/9808060].
}

\lref\tseytRRRR{  A.~A.~Tseytlin,
  ``R**4 terms in 11 dimensions and conformal anomaly of (2,0) theory,''
  Nucl.\ Phys.\ B {\bf 584}, 233 (2000)
  [arXiv:hep-th/0005072].
}

\lref\antRRRR{  I.~Antoniadis, S.~Ferrara, R.~Minasian and
K.~S.~Narain,
  ``R**4 couplings in M- and type II theories on Calabi-Yau spaces,''
  Nucl.\ Phys.\ B {\bf 507}, 571 (1997)
  [arXiv:hep-th/9707013].
 }
  
\lref\WittenMfive{ E.~Witten,
  ``Five-brane effective action in M-theory,''
  J.\ Geom.\ Phys.\  {\bf 22}, 103 (1997)
  [arXiv:hep-th/9610234].
}

\lref\wittenAdS{ E.~Witten,
  ``Anti-de Sitter space and holography,''
  Adv.\ Theor.\ Math.\ Phys.\  {\bf 2}, 253 (1998)
  [arXiv:hep-th/9802150].
  }

\lref\NojiriDH{
  S.~Nojiri and S.~D.~Odintsov,
   ``Conformal anomaly for dilaton coupled theories from AdS/CFT
  correspondence,''
  Phys.\ Lett.\ B {\bf 444}, 92 (1998)
  [arXiv:hep-th/9810008].
}

\lref\brownhen{  J.~D.~Brown and M.~Henneaux,
 ``Central Charges In The Canonical Realization Of Asymptotic Symmetries: An
  Example From Three-Dimensional Gravity,''
  Commun.\ Math.\ Phys.\  {\bf 104}, 207 (1986).
  }

 \lref\wald{
  R.~M.~Wald,
  ``Black hole entropy is the Noether charge,''
  Phys.\ Rev.\ D {\bf 48}, 3427 (1993)
  [arXiv:gr-qc/9307038].
R.~Wald, Phys.\ Rev.\ D {\bf 48} R3427 (1993);
   V.~Iyer and R.~M.~Wald,
  ``Some properties of Noether charge and a proposal for dynamical black hole
  entropy,''
  Phys.\ Rev.\ D {\bf 50}, 846 (1994)
  [arXiv:gr-qc/9403028].
 ``A Comparison of Noether charge and Euclidean methods for computing the
  entropy of stationary black holes,''
  Phys.\ Rev.\ D {\bf 52}, 4430 (1995)
  [arXiv:gr-qc/9503052].
}

\lref\Sen{  A.~Sen,
  ``How does a fundamental string stretch its horizon?,''
  JHEP {\bf 0505}, 059 (2005)
  [arXiv:hep-th/0411255];
   ``Black holes, elementary strings and holomorphic anomaly,''
  arXiv:hep-th/0502126;
   ``Stretching the horizon of a higher dimensional small black hole,''
  arXiv:hep-th/0505122; ``Black hole entropy function and the attractor mechanism in higher
  derivative gravity,''
  arXiv:hep-th/0506177; ``Entropy Function for Heterotic Black Holes,''
  arXiv:hep-th/0508042.
  }

\lref\saidasoda{
  H.~Saida and J.~Soda,
  ``Statistical entropy of BTZ black hole in higher curvature gravity,''
  Phys.\ Lett.\ B {\bf 471}, 358 (2000)
  [arXiv:gr-qc/9909061].
}

 \lref\DDMP{
 A.~Dabholkar, F.~Denef, G.~W.~Moore and B.~Pioline,
  ``Exact and asymptotic degeneracies of small black holes,''
  JHEP {\bf 0508}, 021 (2005)
  [arXiv:hep-th/0502157];
   ``Precision counting of small black holes,''
  JHEP {\bf 0510}, 096 (2005)
  [arXiv:hep-th/0507014].
}

\lref\curvcorr{A.~Dabholkar, ``Exact counting of black hole
microstates", [arXiv:hep-th/0409148],
A.~Dabholkar, R.~Kallosh and A.~Maloney, ``A stringy cloak for a
classical singularity'', JHEP {\bf 0412}, 059 (2004),
[arXiv:hep-th/0410076].
}

\lref\BMPV{ J.~C.~Breckenridge, R.~C.~Myers, A.~W.~Peet and
C.~Vafa, ``D-branes and spinning black holes'', Phys.\ Lett.\ B
{\bf 391}, 93 (1997); [arXiv:hep-th/9602065].
}

\lref\CardosoFP{
  G.~Lopes Cardoso, B.~de Wit and T.~Mohaupt,
   ``Macroscopic entropy formulae and non-holomorphic corrections for
  supersymmetric black holes'',
  Nucl.\ Phys.\ B {\bf 567}, 87 (2000)
  [arXiv:hep-th/9906094];
  ``Deviations from the area law for supersymmetric black holes'',
  Fortsch.\ Phys.\  {\bf 48}, 49 (2000)
  [arXiv:hep-th/9904005];
  ``Corrections to macroscopic supersymmetric black-hole entropy'',
  Phys.\ Lett.\ B {\bf 451}, 309 (1999)
  [arXiv:hep-th/9812082].
}

\lref\hensken{  M.~Henningson and K.~Skenderis,
  ``The holographic Weyl anomaly,''
  JHEP {\bf 9807}, 023 (1998)
  [arXiv:hep-th/9806087].
  }

\lref\deHaroXN{
  S.~de Haro, S.~N.~Solodukhin and K.~Skenderis,
  ``Holographic reconstruction of spacetime and renormalization in the  AdS/CFT
  correspondence,''
  Commun.\ Math.\ Phys.\  {\bf 217}, 595 (2001)
  [arXiv:hep-th/0002230].
}

\lref\PapadimitriouII{
  I.~Papadimitriou and K.~Skenderis,
  ``Thermodynamics of asymptotically locally AdS spacetimes,''
  arXiv:hep-th/0505190.
}

\lref\HollandsWT{
  S.~Hollands, A.~Ishibashi and D.~Marolf,
  ``Comparison between various notions of conserved charges in asymptotically
  AdS-spacetimes,''
  Class.\ Quant.\ Grav.\  {\bf 22}, 2881 (2005)
  [arXiv:hep-th/0503045]; ``Counter-term charges generate bulk symmetries,''
  arXiv:hep-th/0503105.
}

\lref\AlvarezGaumeIG{
  L.~Alvarez-Gaume and E.~Witten,
  ``Gravitational Anomalies,''
  Nucl.\ Phys.\ B {\bf 234}, 269 (1984).
}

\lref\GinspargQN{
  P.~H.~Ginsparg,
  ``Applications Of Topological And Differential Geometric Methods To Anomalies
  In Quantum Field Theory,''
HUTP-85/A056
{\it To appear in Proc. of 16th GIFT Seminar on Theoretical
Physics, Jaca, Spain, Jun 3-7, 1985} }

\lref\BardeenPM{
  W.~A.~Bardeen and B.~Zumino,
  ``Consistent And Covariant Anomalies In Gauge And Gravitational Theories,''
  Nucl.\ Phys.\ B {\bf 244}, 421 (1984).
}

\lref\AlvarezGaumeDR{
  L.~Alvarez-Gaume and P.~H.~Ginsparg,
  ``The Structure Of Gauge And Gravitational Anomalies,''
  Annals Phys.\  {\bf 161}, 423 (1985)
  [Erratum-ibid.\  {\bf 171}, 233 (1986)].
}

\lref\BrownBR{
  J.~D.~Brown and J.~W.~.~York,
  ``Quasilocal energy and conserved charges derived from the gravitational
  Phys.\ Rev.\ D {\bf 47}, 1407 (1993).
}

\lref\fef{C. Fefferman and C.R. Graham, ``Conformal Invariants",
in {\it Elie Cartan et les Math\'{e}matiques d'aujourd'hui}
(Ast\'{e}risque, 1985) 95.}

\lref\BanadosWN{
  M.~Banados, C.~Teitelboim and J.~Zanelli,
  ``The Black hole in three-dimensional space-time,''
  Phys.\ Rev.\ Lett.\  {\bf 69}, 1849 (1992)
  [arXiv:hep-th/9204099];  M.~Banados, M.~Henneaux, C.~Teitelboim and J.~Zanelli,
  ``Geometry of the (2+1) black hole,''
  Phys.\ Rev.\ D {\bf 48}, 1506 (1993)
  [arXiv:gr-qc/9302012].
}

\lref\DeserWH{
  S.~Deser, R.~Jackiw and S.~Templeton,
  ``Topologically Massive Gauge Theories,''
  Annals Phys.\  {\bf 140}, 372 (1982)
  [Erratum-ibid.\  {\bf 185}, 406.1988\ APNYA,281,409
  (1988\ APNYA,281,409-449.2000)]; ``Three-Dimensional Massive Gauge Theories,''
  Phys.\ Rev.\ Lett.\  {\bf 48}, 975 (1982).
}

\lref\MooreFG{
  G.~W.~Moore,
  ``Les Houches lectures on strings and arithmetic,''
  arXiv:hep-th/0401049.
}

\lref\ElitzurNR{
  S.~Elitzur, G.~W.~Moore, A.~Schwimmer and N.~Seiberg,
  ``Remarks On The Canonical Quantization Of The Chern-Simons-Witten Theory,''
  Nucl.\ Phys.\ B {\bf 326}, 108 (1989).
}

\lref\AharonyTI{
  O.~Aharony, S.~S.~Gubser, J.~M.~Maldacena, H.~Ooguri and Y.~Oz,
  ``Large N field theories, string theory and gravity,''
  Phys.\ Rept.\  {\bf 323}, 183 (2000)
  [arXiv:hep-th/9905111].
}

\lref\PiolineNI{
  B.~Pioline,
  ``Lectures on on black holes, topological strings and quantum attractors,''
  arXiv:hep-th/0607227.
}

\lref\DavidWN{
  J.~R.~David, G.~Mandal and S.~R.~Wadia,
  ``Microscopic formulation of black holes in string theory,''
  Phys.\ Rept.\  {\bf 369}, 549 (2002)
  [arXiv:hep-th/0203048].
}

\lref\EmparanPM{
  R.~Emparan, C.~V.~Johnson and R.~C.~Myers,
  ``Surface terms as counterterms in the AdS/CFT correspondence,''
  Phys.\ Rev.\ D {\bf 60}, 104001 (1999)
  [arXiv:hep-th/9903238].
}

\lref\CarlipGC{
  S.~Carlip and C.~Teitelboim,
   ``Aspects Of Black Hole Quantum Mechanics And Thermodynamics In
  (2+1)-Dimensions,''
  Phys.\ Rev.\ D {\bf 51}, 622 (1995)
  [arXiv:gr-qc/9405070].
}

\lref\CallanSA{
  C.~G.~.~Callan and J.~A.~Harvey,
  ``Anomalies And Fermion Zero Modes On Strings And Domain Walls,''
  Nucl.\ Phys.\ B {\bf 250}, 427 (1985).
}

\lref\CveticXH{
  M.~Cvetic and F.~Larsen,
  ``Near horizon geometry of rotating black holes in five dimensions,''
  Nucl.\ Phys.\ B {\bf 531}, 239 (1998)
  [arXiv:hep-th/9805097].
}

\lref\MinasianQN{
  R.~Minasian, G.~W.~Moore and D.~Tsimpis,
  ``Calabi-Yau black holes and (0,4) sigma models,''
  Commun.\ Math.\ Phys.\  {\bf 209}, 325 (2000)
  [arXiv:hep-th/9904217].
}

\lref\Bott{R.~Bott and A.S.~Cattaneo, ``Integral invariant of
3-manifolds," [arxiv:dg-ga/9710001].}

\lref\GoldsteinHQ{
  K.~Goldstein, N.~Iizuka, R.~P.~Jena and S.~P.~Trivedi,
  ``Non-supersymmetric attractors,''
  Phys.\ Rev.\ D {\bf 72}, 124021 (2005)
  [arXiv:hep-th/0507096].
}

\lref\MaldacenaRE{
  J.~M.~Maldacena,
  ``The large N limit of superconformal field theories and supergravity,''
  Adv.\ Theor.\ Math.\ Phys.\  {\bf 2}, 231 (1998)
  [Int.\ J.\ Theor.\ Phys.\  {\bf 38}, 1113 (1999)]
  [arXiv:hep-th/9711200].
}

\lref\PolchinskiRQ{
  J.~Polchinski,
  ``String theory''
}

\lref\KawaiJK{
  T.~Kawai, Y.~Yamada and S.~K.~Yang,
  ``Elliptic Genera And N=2 Superconformal Field Theory,''
  Nucl.\ Phys.\ B {\bf 414}, 191 (1994)
  [arXiv:hep-th/9306096].
}

\lref\LuninIZ{
  O.~Lunin, J.~M.~Maldacena and L.~Maoz,
  ``Gravity solutions for the D1-D5 system with angular momentum,''
  arXiv:hep-th/0212210.
}

\lref\SolodukhinAH{
  S.~N.~Solodukhin,
  ``Holography with gravitational Chern-Simons,''
  Phys.\ Rev.\ D {\bf 74}, 024015 (2006)
  [arXiv:hep-th/0509148]; ``Holographic description of gravitational anomalies,''
  JHEP {\bf 0607}, 003 (2006)
  [arXiv:hep-th/0512216].
}

\lref\SahooRP{
  B.~Sahoo and A.~Sen,
   ``Higher derivative corrections to non-supersymmetric extremal black holes in
  N = 2 supergravity,''
  arXiv:hep-th/0603149.
}

\lref\EmparanIT{
  R.~Emparan and G.~T.~Horowitz,
  ``Microstates of a neutral black hole in M theory,''
  arXiv:hep-th/0607023.
}

\lref\MohauptMJ{
  T.~Mohaupt,
  ``Black hole entropy, special geometry and strings,''
  Fortsch.\ Phys.\  {\bf 49}, 3 (2001)
  [arXiv:hep-th/0007195].
}

\lref\PeetHN{
  A.~W.~Peet,
  ``TASI lectures on black holes in string theory,''
  arXiv:hep-th/0008241.
  }

\lref\MathurAI{
  S.~D.~Mathur,
  ``The quantum structure of black holes,''
  Class.\ Quant.\ Grav.\  {\bf 23}, R115 (2006)
  [arXiv:hep-th/0510180].
}

\lref\MaldacenaKY{
  J.~M.~Maldacena,
  ``Black holes in string theory,''
  arXiv:hep-th/9607235.
}

\lref\iosif{I.~Bena, Frascati 2006 lectures}

\lref\ImbimboBJ{
  C.~Imbimbo, A.~Schwimmer, S.~Theisen and S.~Yankielowicz,
  ``Diffeomorphisms and holographic anomalies,''
  Class.\ Quant.\ Grav.\  {\bf 17}, 1129 (2000)
  [arXiv:hep-th/9910267].
}


\Title{\vbox{\baselineskip12pt
}} {\vbox{\centerline {Lectures on black holes and the
AdS$_3$/CFT$_2$  correspondence}} }
\centerline{Per Kraus\foot{pkraus@ucla.edu}}

\bigskip
\centerline{$$\it{Department of Physics and Astronomy,
UCLA,}}\centerline{\it{ Los Angeles, CA 90095-1547, USA.}}

\baselineskip15pt

\vskip .3in

\centerline{\bf Abstract\foot{Presented at the Winter School on
the Attractor Mechanism (Frascati, March 20-24, 2006)}}

We present a detailed discussion of AdS$_3$ black holes and their
connection to two-dimensional conformal field theories via the
AdS/CFT correspondence.  Our emphasis is on deriving refined
versions of black hole partition functions, that include the
effect of higher derivative terms in the spacetime action as well
as non-perturbative effects.  We include background material on
gravity in AdS$_3$, in the context of holographic renormalization.

\Date{September, 2006}


\listtoc \writetoc

\newsec{Introduction}

The fact that string theory is able to provide a successful
microscopic description of certain black holes provides strong
evidence that it is a consistent theory of quantum gravity.
Correctly reproducing the Bekenstein-Hawking entropy formula
$S=A/4G$ from an explicit sum over states indicates that the right
microscopic degrees of freedom have been identified.   Since
string theory also reduces to conventional general relativity
(coupled to matter) at low energy, it seems to provide us with a
coherent theory encompassing both the microscopic and macroscopic
regimes. Needless to say, however, there is still much to be
learned about the full implications of string theory for quantum
gravity.

One approach to deepening our understanding is to examine the
string theory description of black holes with improved precision.
This program has been highly fruitful so far.   The earliest
successful black hole entropy matches, following \StromingerSH,
appeared somewhat miraculous, the emergence of the
Bekenstein-Hawking formula from the microscopic side not becoming
apparent until all the last numerical factors were accounted for.
The precise agreement seemed even more astonishing once additional
features like rotation and non-extremality were included.   It was
eventually understood that the essential ingredients on the two
sides are the near horizon AdS region of the black hole geometry,
and the low energy CFT describing the underlying branes, and this
(among other observations) led to the celebrated AdS/CFT
correspondence \MaldacenaRE.   This improved understanding largely
demystifies the nature of the entropy matching, as we'll discuss
in these lectures.  One of our goals here will be to show how just
a few basic features, like the existence of a near horizon AdS
region with the appropriate symmetries, is enough to make the
agreement manifest, even in rather complicated contexts, and
including incorporating subleading corrections to the area law
formula.

A survey of the examples in which there is a precise microscopic
accounting of black entropy reveals the near ubiquitous appearance
of a near horizon AdS$_3$ factor (possibly after a suitable
duality transformation).\foot{For a recent example without such an
AdS$_3$ factor see \EmparanIT.  But note that in this example the
microscopic counting is not under complete control, and
interestingly still involves relating the system to another system
which {\it does} have an AdS$_3$ region.}  In these examples the
dual theory is a two-dimensional CFT,  for which there are
powerful results constraining the spectrum of states.    By
contrast, in other examples such as AdS$_5$ black holes, it has so
far only been possible to compute the entropy up to at best
numerical factors.    For this reason, here we will be focussing
on  AdS$_3$ examples.

The AdS$_3$/CFT$_2$ correspondence can be stated as an equivalence
between partition functions
\eqn\aa{Z_{AdS} = Z_{CFT}~.}
The connection with black hole entropy arises when we examine this
relation in the high energy regime, where the left hand side is
dominated by an asymptotically AdS$_3$ black hole: the BTZ black
hole \BanadosWN.   General properties of conformal field theories
imply that the asymptotic density of states will agree between the
two sides, as we'll discuss in what follows

A more ambitious goal is to try to demonstrate {\it exact}
agreement in \aa.   On the gravitational side this will involve
incorporating many new contributions beyond that of a single large
black hole.  One way to think of defining $Z_{AdS}$ is as a
Euclidean path integral.  At finite temperature the contributing
Euclidean geometries should have a boundary that is a
two-dimensional torus, to match with the standard finite
temperature description of the boundary CFT.   A typical bulk
geometry that is thereby included is one whose topology is  a
three-dimensional solid torus.  Such a geometry appears in the
path integral weighted by its Euclidean action, which includes (if
we are trying to be exact)  contributions from an infinite series
of higher derivative terms in the spacetime Lagrangian.  That is
not all though, since we also need to include all possible
excitations on top of the geometry, allowing for particle, string,
and brane states that can wind around the solid torus.  After all
these contributions have been taken into account one can hope to
match to the exact CFT partition function.

In these notes we will discuss to what extent this program can be
carried out.  This will involve a careful study of gravity in
asymptotically AdS$_3$ spacetimes, and its string theory
realization.   We organize our presentation by starting with a
fairly generic setup and then becoming progressively more
specific. As we'll see, once we start adding more structure, like
supersymmetry, to the problem, and refine our definitions of the
partition functions in \aa, it is possible to go a significant
distance in demonstrating exact agreement between the
gravitational and CFT descriptions.

Let us describe the outcome of our analysis in a bit more detail.
In a two-dimensional CFT we have independent temperatures for the
left and right movers; we label the inverse left(right) moving
temperature as $\tau \sim 1/T_L$ ($\taub \sim 1/T_R$).  Further,
in the CFT there is a spectrum of left and right moving conserved
charges, and we can turn on chemical potentials for these charges,
$z_I$ and $\tilde{z}_I$.  Allowing for nonzero potentials lets us
study charged black holes.   To study black hole entropy we are
interested in the high temperature behavior of the partition
function, and we will see that it has the structure
\eqn\ab{\eqalign{\ln Z &= {i\pi \over \tau }({c\over 12} -2
C^{IJ}z_I z_J) - {i\pi \over \taub} ({\ct \over 12} -2
\tilde{C}^{IJ}\zt_i \zt_J)\cr &\quad+ {\rm
exponentially~suppressed~terms}~. }}
Here $c$ and $\ct$ are the left and right moving central charges,
and  $C^{IJ}$ and $\tilde{C}^{IJ}$ are matrices appearing in the
CFT current algebra.   On the gravity side, if we use the
two-derivative approximation to the spacetime action, and discard
the exponentially small terms, we will reproduce the area law for
the entropy of a general rotating, charged black hole.  But we can
go considerably further:  the parameters $c$, etc., can be
computed {\it exactly} by relating them to anomalies.  The
corrections to these parameters encode the effect of higher
derivative terms in the spacetime action, and lead to corrections
to the area law. Indeed, by transforming \ab\ into an expression
for the degeneracy as a function of charges (i.e. relating the
canonical ensemble to the microcanonical ensemble via a Laplace
transform) we deduce a series of $1/Q$ corrections to the
degeneracy, as in \DDMP.  The suppressed terms in the second line
of \ab\ will arise, in the gravitational description, from
including fluctuations around black hole geometries, and from
summing over inequivalent black holes.

As we proceed, it will become clear that terms on the first and
second lines of \ab\ are of a rather different nature.  The top
line can be established on general grounds, using the relation to
anomalies.  In particular, this can be achieved even for non-BPS
and nonextremal black hole, and so a class of  area law
corrections for such black holes are under excellent
control.\foot{Here we mean that we can consider non-supersymmetric
black hole solutions to an underlying supersymmetric theory.
Corrections can be computed also for theories with no underlying
supersymmetry (as we'll discuss in section 2) but explicit
knowledge of the full Lagrangian is needed in these cases.} Also,
our method of derivation will make it manifest that the black hole
and CFT entropies agree in these cases. The second line of \ab,
however, is much more context dependent, and further can only be
computed explicitly when we define $Z$ to be a supersymmetric
partition function (an index).

In section 2 we begin with pure gravity in asymptotically AdS$_3$
spacetimes.  We review how to properly  define the gravitational
action by including boundary terms, and then review the
construction of the boundary stress tensor dual to the stress
tensor of the CFT.  In two-derivative gravity, this stress tensor
obeys a Virasoro algebra with the  central charge of Brown and
Henneaux \brownhen.  We then generalize to higher derivative
theories of gravity, and show how to obtain the generalized
central charge. It turns out that the central charge can be found
by a simple extremization principle. With these results in hand,
we turn to computing the entropy of BTZ black holes in general
higher derivative theories. A crucial role here is played by the
construction of BTZ as a quotient of AdS, and its relation to a
thermal AdS geometry via a modular transformation. This analysis
will establish the agreement between the black hole and CFT
entropies once the central charges have been shown to agree.

In section 3 we add gauge fields into the mix and show how these
are dual to currents in the boundary CFT.   A central role is
played by bulk Chern-Simons terms, since these turn out to
completely determine the currents.   Turning on flat connections
for our gauge fields allows us to incorporate charged black holes.
We then discuss the role of a Chern-Simons term for the
gravitational field, and show how it is used to deduce the
difference between the central charges of the left and right
moving sectors of the CFT.

The two specific string theory constructions that we'll consider
are reviewed in section 4: the D1-D5 system giving rise to
five-dimensional black holes with near horizon geometry
AdS$_3\times S^3$, and wrapped M5-branes yielding four-dimensional
black holes with near horizon geometry AdS$_3 \times S^2$.  To
read off the exact central charges for these systems we will use a
combination of anomalies and supersymmetry.  In particular, this
will allow us to derive the exact corrections to the classical
central charges, and hence derive a class of corrections to the
black hole area law.   The main emphasis is in showing how these
exact results can be obtained even without knowing the explicit
form of all higher derivative terms in the spacetime action. A
nice application of this formalism is to small black holes dual to
fundamental heterotic strings, and we will show how to derive the
worldsheet central charges from gravity.

In sections 5-8 we turn to the computation of the full partition
function from the gravitational point of view.  In order to have a
chance of making an exact computation we focus on the elliptic
genus, which is a particular partition function invariant under
smooth deformations of the theory, due to bose-fermi
cancellations.  We review its main properties in section 5, and
then show in sections 6 and 7 how these properties emerge in the
gravity description.   Much of our discussion will follow the work
of Dijkgraaf et. al. \farey\ on the ``Farey tail" description of
the elliptic genus for the D1-D5 system.  This gives a beautiful
example of the matching between the CFT and gravity versions of
the elliptic genus, including the effects of summing over
geometries.    Section 8 shows how to incorporate the effects of
BPS excitations of top of the background geometries being summed
over in the path integral.  These include both supergravity
fluctuations from Kaluza-Klein reduction, as well as
non-perturbative brane states.   In our brief discussion of the
latter, following \GaiottoNS\ we note how the appearance of both
branes and anti-brane BPS states leads to the OSV formula
\OoguriZV\ relating the AdS partition function  to that of the
topological string.

\vskip.3cm \noindent {\it General references:}

In these lectures we  focus on one particular aspect of black hole
physics, namely the computation of the entropy/partition function
of AdS$_3$ black holes, and   our presentation is based mainly on
\refs{\KrausVZ,\KrausNB}.  There are of course many other major
issues that we will not touch on substantially, such as the
information paradox, the black hole singularity, and black holes
in other dimensions. There are a number of excellent pedagogical
treatments of various aspects of black holes in string theory that
complement the material discussed here.  An incomplete list is
\refs{\MaldacenaKY,\AharonyTI,\MohauptMJ,\PeetHN,\DavidWN,
\MathurAI,\PiolineNI }.

\newsec{Gravity in asymptotically AdS$_3$ spacetimes}

\subsec{Action and stress tensor}

In this section we  consider  pure gravity in three dimensions in
the presence of a negative cosmological constant.    This theory
is  described by an Einstein-Hilbert action supplemented by
boundary terms
\eqn\ba{I=  {1 \over 16\pi G } \int \! d^3 x \sqrt{g} \,(R-{2
\over \ell^2}) +I_{\rm bndy}~.}
The need for, and explicit form of, the boundary terms in the
action will become clear as we proceed. We work in Euclidean
signature, and follow the curvature conventions of Misner, Thorne,
and Wheeler.

One solution of the equations of motion is AdS$_3$,
\eqn\bb{ ds^2 = (1+r^2/\ell^2)dt^2 +{dr^2 \over 1+r^2/\ell^2}+r^2 d\phi^2~.}
AdS$_3$ is homogeneous space of  constant negative curvature.  It
has maximal symmetry, the isometry group being $SL(2,\IC) \cong
SL(2,\IR)_L \times SL(2,\IR)_R$ as will be reviewed later.    The
metric \bb\ is written in so-called global coordinates that cover
the entire manifold.   AdS$_3$ will play the role of the vacuum of
our theory, in that it has the lowest mass of any solution.  In
fact, we'll see that it is natural to assign it the negative mass
$M= -{\ell \over 8 G}$.

A more general one-parameter family of solutions is the
non-rotating  BTZ black hole \BanadosWN,
\eqn\bc{ ds^2 = {(r^2- r_+^2) \over \ell^2}dt^2 + {\ell^2  \over
(r^2-r_+^2)}dr^2 + r^2 d\phi^2~.}
After rotating to Lorentzian signature it is evident that  this
describes a black with event horizon at $r=r_+$, and
Bekenstein-Hawking entropy
\eqn\bez{  S  = {A \over 4G } = {\pi r_+ \over 2 G}~.}
Note that if we set $r_+^2 = -\ell^2$ we recover \bb.   The black
hole solution \bc\ can be further generalized by adding charge and
rotation; we will have much more to say about this.

By examining the large $r$ behavior, it is apparent that the
solution \bc\ asymptotically approaches AdS$_3$.  We now want to
state the precise conditions under which a metric can be said to
be asymptotically  AdS$_3$.   This is a standard type of question
in general relativity, and can be approached from different
viewpoints.   Our focus will be on demanding the existence of a
well defined action and variational principle.  A motivation for
this from the point of view of AdS/CFT is that the action takes on
a well defined meaning as giving, in a suitable semiclassical
limit,  the partition function of the CFT; indeed this is
essentially the fundamental definition of the AdS/CFT
correspondence.   We would also like to include as large a class
of metrics as possible.   Furthermore, we have the freedom  to
adjust the boundary terms in \ba\ to make the action finite and
stationary when the Einstein equations are satisfied.

To analyze this problem it is convenient to work in coordinates
where the metric takes the form (Gaussian normal coordinates)
\eqn\bfz{ds^2 = d\eta^2 + g_{ij} dx^i dx^j~.}
Here $g_{ij}$ is an arbitrary function of $x^i$ ($i=1,2)$ and the
radial coordinate $\eta$. The allowed values of  $\eta$ are
unbounded from above, although there may be a minimal value
imposed by smoothness considerations.    Now, it is apparent that
the action written in \ba\ will diverge due to the large $\eta$
integration, and so we regulate the integral by imposing a cutoff
at some fixed value of $\eta$, which we eventually hope to take to
infinity.

In terms of \bfz\ the bulk term in the action \ba\ appears as,
after an integration by parts,
\eqn\bg{I_{EH}  = {1 \over 16\pi G } \int
\! d^2x \,d\eta \sqrt{g} \left(R^{(2)}+ (\Tr K)^2- \Tr K^2-2\Lambda
\right) - {1 \over 8 \pi G}\int_{\p {\cal M}} \! d^2 x \sqrt{g}\,
\Tr K~,}
where $R^{(2)}$ is the Ricci scalar associated with $g_{ij}$.
$K$ is the extrinsic curvature, defined as
\eqn\bh{ K_{ij} = \half \p_\eta  g_{ij}~.}
All indices are raised and lowered by $g_{ij}$ and its inverse.

The variation of the boundary term contains a contribution $\delta
\p_\eta g_{ij}$.   This term spoils a variational principle in
which we hold fixed the induced metric on $\p {\cal M}$, but not
its  normal derivative.  This is rectified by adding to the action
the Gibbons-Hawking term
\eqn\be{I_{GH} = {1 \over 8 \pi G}\int_{\p {\cal M}} \! d^2 x
\sqrt{g}\, \Tr K~. }

We now consider the variation of the action with respect to
$g_{ij}$.  The variation will consist of two terms: a bulk piece
that vanishes when the equations of motion are satisfied, and a
boundary piece.  Assuming that the equations of motion are
satisfied, a simple computation gives
\eqn\bff{\delta (I_{EH}+I_{GH}) = -{1 \over 16\pi G} \int_{\p
{\cal M}}\! d^dx \sqrt{g} \, (K^{ij} - \Tr K g^{ij})\delta
g_{ij}~.}
The boundary stress tensor (which in the AdS/CFT correspondence
is dual to the CFT stress tensor)   is defined in terms of the
variation as
\eqn\bfa{ \delta I = \half \int_{\p{\cal M}} \! d^2x \sqrt{g}\,
T^{ij} \delta g_{ij}~,}
 and so we have at this stage
\eqn\bg{ T^{ij} = -{1 \over 8\pi G} (K^{ij} - \Tr K g^{ij})~,}
which is a result derived by Brown and York \BrownBR. Although
we derived this result in the coordinate system \ba, the result
\bg\ is valid in any coordinate system, where $g_{ij}$ is the
induced metric on the boundary, and $K_{ij}$ is the extrinsic
curvature.

We now incorporate the specific features of asymptotically AdS
spacetimes, which will require the addition of a second boundary
term. From the basic solutions \bb-\bc\ it is evident that we
should allow metrics that grow as $r^2$ at infinity. Translating
to the $\eta$ coordinate, this implies a growth $e^{2\eta/\ell}$.
By studying the Einstein equations one finds that the general
solution has subleading terms down by powers of $e^{-2\eta/\ell}$.
We therefore write a ``Fefferman-Graham expansion" \fef\  for the
metric as
\eqn\bh{g_{ij}  = e^{2\eta/\ell}g^{(0)}_{ij} + g^{(2)}_{ij} +
\ldots~. }
 Omitted terms fall off at least as $e^{-\eta/\ell}$.
$g^{(0)}_{ij}$ is the ``conformal boundary metric"; it is clearly
defined only up to Weyl transformations induced by a redefinition
of $\eta$.     It is this metric that we wish to identify with the
metric of the boundary CFT.

Given  $g^{(0)}_{ij}$,  the subleading terms in the expansion \bh\
are found by solving Einstein's equations.    Here we just note
the following important relation that arises (see e.g. \deHaroXN):
\eqn\bha{    \Tr (g^{(2)})= \half \ell^2 R^{(0)}~,}
where indices are lowered and raised with $\go_{ij}$ and its inverse
$g^{(0) ij}$.

Upon removal of the large $\eta$ regulator, it is clear the
$g^{(0)}_{ij}$ plays the role of the boundary metric, and so it is
natural to seek a variational principle in which $g^{(0)}_{ij}$ is
held fixed, while the subleading parts of \bh\ are allowed to
vary.\foot{In higher dimensional AdS spacetimes a finite number of
subleading terms are determined {\it algebraically} in terms of
$g^{(0)}_{ ij}$, and are therefore also kept fixed.  See, e.g.,
\deHaroXN} However, our action $I_{EH}+I_{GH}$ fails on two
counts. First, using \bff\ it is not hard to check that the
variation of $g^{(2)}_{ij}$ appears explicitly, and second that
\bff\ diverges in the large $\eta$ limit.   Both of these problems
are solved by adding to the action the ``counterterm"
\refs{\hensken,\BalasubramanianRE}
\eqn\bi{ I_{ct}= -{1 \over 8\pi G \ell} \int_{\p {\cal M}} \! d^2x \sqrt{g}~.}
Once this is included,  it is straightforward  to check that the
on-shell variation of the action takes the form
\eqn\bj{ \delta I = \half \int  \! d^2x \sqrt{g^{(0)}}\,
T^{ij} \delta g^{(0)}_{ij}~,}
with
\eqn\bja{T_{ij} = {1 \over 8\pi G \ell}
\left(g^{(2)}_{ij}-\Tr (g^{(2)} ) g^{(0)}_{ij}\right)
~.}
This is our AdS$_3$ stress tensor. The stress tensors for higher
dimensional spacetimes can be found in the literature
\refs{\BalasubramanianRE,\EmparanPM}, and additional related work
appears in
\refs{\KrausDI,\deHaroXN,\PapadimitriouII,\HollandsWT,\NojiriDH}.

Note that the stress tensor has a nonzero trace \hensken,
\eqn\bk{ \Tr (T) = -{1 \over 8\pi G\ell}  \Tr(g^{(2)}) = -{\ell
\over 16\pi G } R^{(0)}~,}
where we used \bha.    This is the Weyl anomaly.  In fact, the
stress tensor defined here obeys all the properties of a stress
tensor in CFT, and we can thereby read off the central charge by
comparing to the standard form of the Weyl anomaly, $\Tr(T)  = -
{c \over 24 \pi} R$. This gives the central charge originally
derived by Brown and Henneaux \brownhen,
\eqn\bl{  c= {3\ell \over 2G}~.}

In the absence of the Weyl anomaly we can  think of $g^{(0)}$ as
specifying a conformal class of metrics, and the action is
independent of the particular representative we choose.   But when
the Weyl anomaly is nonvanishing we need to choose a specific
representative.   Another way to understand the Weyl anomaly is
that although we succeeded in making the variation \bj\ finite, it
is not hard to check that the action itself can suffer from a
divergence linear in $\eta$. To cancel this divergence we are
forced to add another counterterm that depends explicitly (and
linearly)  on our large $\eta$ cutoff.  The Weyl anoomaly can then
be read off from the transformation of this term under a shift in
the cutoff.

\subsec{Virasoro generators}

To simplify the discussion, it is now convenient to  take
$g^{(0)}_{ij}$ to be a flat metric on the cylinder and to work in
complex coordinates.   We thus take $g^{(0)}_{ij}dx^i dx^j = dw
d\wb  $ with $w  \cong w+2\pi$.   When we write $w =\sigma_1 +i
\sigma_2$ we'll think of $\sigma_2$ as the imaginary time
direction.   The stress tensor now has components
\eqn\bm{ T_{ww} = {1 \over 8\pi G \ell} g^{(2)}_{ww}~,\quad T_{\wb\wb} = {1 \over 8\pi G \ell} g^{(2)}_{\wb\wb}~.}
$T_{ww}$ ($T_{\wb\wb}$) is holomorphic (anti-holomorphic)  as a
consequence of the Einstein equations.

The Virasoro generators are defined in the usual fashion  as
contour integrals
\eqn\bn{\eqalign{ L_n - {c\over 24} \delta_{n,0}& = \oint \! dw~ e^{-in w} T_{ww} \cr
\Lt_n - {\ct\over 24} \delta_{n,0}& = \oint \! d\wb~ e^{in \wb} T_{\wb\wb}~.}}
Looking ahead, we have allowed for an independent  rightmoving
central charge $\ct$, although at this stage $c=\ct$.     The
generators obey the Virasoro algebra
\eqn\bo{ [L_m,L_n] = (m-n)L_{m+n} +{c\over
12}(m^3-m)\delta_{m+n}~,}
and likewise for the $\Lt_n$.   To establish this one  studies the
transformation of the stress tensor under the coordinate
transformations that preserve the form of $g^{(0)}$.   The
infinitesimal transformation law is then used to derive  the
algebra \bo.

Mass and angular momentum in AdS$_3$ are related to the  Virasoro
charges as
\eqn\bp{ L_0 -{c\over 24} = \half (M\ell - J)~,\quad   \Lt_0
-{\ct\over 24} = \half (M\ell + J)~.}

As a simple example consider the BTZ metric \bc.     We find
$g^{(2)}_{ww}=g^{(2)}_{\wb\wb} = r_+^2/4$, and hence
\eqn\bq{ L_0 = \Lt_0 = {\ell \over 16 G}(1+{r_+^2 \over\ell^2})~,}
or
\eqn\br{ M = {r_+^2 \over 8 G \ell^2}~,\quad J=0~.}
Note that the pure AdS$_3$ metric \bb\ has $L_0  = \Lt_0=0$, which
is simply a consequence of its invariance under the $SL(2,\IR)_L
\times SL(2,\IR)_R $ group of isometries generated by $L_{0,\pm
1}$ and  $\Lt_{0,\pm 1}$.

\subsec{Generalization to higher derivative theories
\ImbimboBJ,\saidasoda,\KrausVZ}

In the preceding we have been working with a two derivative
action.  In the context of string theory, or any other sensible
approach to quantum gravity, this will just be the leading part of
a more general effective action containing  terms with arbitrary
numbers of derivatives. If we are to make precise statements about
such physical quantities as black hole entropy we need a
systematic way of including the effect of higher derivative terms.
For example, we no longer expect the entropy-area relation
$S=A/4G$ to hold in the general case. On the face of it, even if
we knew the explicit form of the action it would seem to be highly
nontrivial to repeat the previous analysis and extract physical
quantities.   But in fact the problem is much easier than it first
appears.

Using the fact that in three  dimensions the Riemann tensor can be
expressed in terms of the Ricci tensor, we may write an arbitrary
higher derivative action as
\eqn\bs{ I = {1 \over 16 \pi G}  \int\! d^3x \sqrt{g}\, {\cal
L}(g^{\mu\nu},\nabla_\mu, R_{\mu\nu} ) + I_{bndy}~.}
In fact, there is one additional  term that can be added, a
gravitational Chern-Simons term related to the possibility of
$c\neq \ct$, that is being suppressed in \bs.  We will come back
to it later.

We now ask how to derive the generalized version of the central
charge formula \bl.     Because AdS$_3$ is maximally symmetric we
know that it will be a solution of our higher derivative theory,
but we need to determine the length scale $\ell$.    To proceed,
we write pure AdS$_3$ in the following coordinates
\eqn\bt{ ds^2 = \ell^2( d\eta^2 + \sinh^2 \eta d\Omega_2^2)~,}
so that $\ell$ only appears as an overall factor.      In \bt\
$\ell$ is of course a constant, but to determine its value it is
useful to consider a {\it local} variation of compact support,
$\ell \rightarrow \ell + \delta \ell (x)$.   When the equations of
motion are satisfied the action should be stationary under such a
variation.   The variation of the action computed around \bt\
takes a very simple form as follows from the fact that all
tensorial  quantities are covariantly constant on AdS$_3$.  A
moments thought then shows that the variation takes the form
\eqn\bu{ \delta I = {1 \over 16\pi G } \int\! d^3x   ~{\p \over \p
\ell} \left( \sqrt{g} {\cal L}\right)\delta \ell(x)~.}
So the equations of motion imply that $\sqrt{g} {\cal L}$  should
be at an extremum with respect to {\it rigid} variations of
$\ell$.     Given the explicit form of ${\cal L}$ we then need
``only" solve an algebraic equation to determine $\ell$.

Now we turn to the determination of the central charge.
Conformal invariance implies the general relation $\Tr (T) =
-{c\over 24 \pi} R^{(0)}$.   Consider \bj\ in the context of an
infinitesimal Weyl transformation, $\delta g^{(0)}_{ij} = 2 \delta
\omega g^{(0)}_{ij}$,  applied to a metric whose boundary is
conformal to $S^2$,
\eqn\bv{ \delta I = {1\over 2} \int\! d^2 x \sqrt{g^{(0)}}\,T^{ij}\delta
g^{(0)}_{ij}   = - {c
\over 24 \pi} \delta \omega \int \! d^2 x\sqrt{g^{(0)}}~ R^{(0)} =-  {c
\over 3} \delta \omega~.}
To extract $c$ we evaluate \bs\ on the metric \bt.    ${\cal L}$
is a constant on this solution since AdS$_3$ is homogeneous, and
so
\eqn\bw{ I = {\ell^3 {\cal L}  \over 4G} \int\! d\eta ~\sinh^2 \eta + I_{bndy}~.}
The integration is divergent at large $\eta$ and so we impose a
cutoff $\eta \leq \eta_{max}$ and write  $\int\! d\eta ~\sinh^2
\eta = -\half \eta_{max} +{1 \over 4}  \sinh( 2\eta_{max})$. Now,
$I_{bndy}$ is built out  of the induced metric on the boundary.
Assuming it is local, we can arrange it to subtract off the
$\sinh(2\eta_{max})$ term but not the linear term.   Indeed, as we
discussed below \bl\ the linear divergence is the Weyl anomaly,
which (like all anomalies) cannot be subtracted by local
counterterms.   So even after adding $I_{bndy}$ the action
diverges as
\eqn\bx{ I_{div} = -{ \ell^3 {\cal L} \over 8G} \eta_{max}~.}
Next, observe that a shift of $\eta_{max}$ implements a Weyl
transformation, $\delta \omega = \delta \eta_{max}$.   We can
therefore equate   \bv\ with the variation of \bx\ to
obtain\foot{Note that Ref. \ImbimboBJ\  also considers the effect
of higher derivatives on conformal anomalies in higher
dimensions.} \refs{\ImbimboBJ,\KrausVZ}
\eqn\by{ c= {3 \ell^3 {\cal L} \over 8G}~.}
Recall that ${\cal L}$ should be evaluated at the extremum of
$\sqrt{g} {\cal L}$.  But given \bt\ we see that $\sqrt{g} {\cal
L} \propto \ell^3 {\cal L} $, so we can equally well say that we
are extremizing $c$.  We have now derived the {\it c-extremization
principle}:  the central charge is obtained by the value of \by\
at its extremum.

Another version of \by\ is also useful.    Extremization  of
$\sqrt{g} {\cal L}$ implies
\eqn\bz{ 3 {\cal L} + 2 \ell^2 {\p {\cal L} \over \p \ell^2} =0~.}
Since all covariant derivatives vanish on  the background we can
ignore them for the purposes of this computation and write ${\cal
L} = {\cal L}(g^{\mu\nu},R_{\mu\nu})$.   Given \bt\ we have  that
$\ell^2$ appears in $g^{\mu\nu}$ but not in $R_{\mu\nu}$.  This
together with the fact that all indices in ${\cal L}$ must be
contracted, implies
\eqn\ca{\ell^2 {\p {\cal L} \over \p \ell^2} =-R_{\mu\nu} {\p {\cal L} \over \p R_{\mu\nu}} =
-{2 \over \ell^2} g_{\mu\nu} {\p {\cal L} \over \p R_{\mu\nu}}~,}
where we also used $R_{\mu\nu} = {2 \over \ell^2} g_{\mu\nu}$ for
\bt.    Using \bz\ and \ca\ we can rewrite \by\ as
\refs{\saidasoda,\KrausVZ}
\eqn\cb{ c = {\ell \over 2G} g_{\mu\nu} {\p {\cal L} \over \p R_{\mu\nu}}~.}
This is the most convenient form for the AdS$_3$ central  charge.
As a quick check, if we return to the action  \ba\ we find ${\p
{\cal L} \over \p R_{\mu\nu}} = g^{\mu\nu}$ and so we recover the
Brown-Henneaux central charge,  $c=3\ell/2G$.

We will now show how to use this result to derive the entropy of a
BTZ black hole in a general higher derivative theory of gravity.

\subsec{Thermal AdS partition function}

The AdS/CFT correspondence is fundamentally a relation  between
partition functions
\eqn\cba{ Z_{AdS}(g^{(0})  =  Z_{CFT}(g^{(0)})~.}
Here we have just indicated the dependence on the metric, although
more generally other data will enter in as well.   Modulo the Weyl
anomaly,  $g^{(0)}$ labels  a conformal class of boundary metrics.

In this section we will consider the case in which $g^{(0)}$ is
the flat metric on a torus of modular parameter $\tau$.  We write
the line element of the boundary in complex coordinates as
$ds^2=dwd\wb$, with
\eqn\cc{ w \cong w+ 2\pi \cong w+ 2\pi \tau~.}
$Z_{CFT}$ can either be evaluated as a path integral on the torus
(assuming that there exists a Lagrangian formulation of the
theory) or in the canonical formulation as
\eqn\cd{ Z_{CFT}(\tau,\taub) = \Tr \left[ e^{2\pi i \tau (L_0-{c
\over 24})} e^{-2\pi i \taub (\Lt_0 - {\ct \over 24})}\right]~.}
If fermions are present in the theory we also need to specify
their periodicities around the two cycles of the torus.   As
written, \cd\ implies anti-periodic boundary conditions around the
time circle; periodic fermions are incorporated by including
$(-1)^F$ in the trace, where $F$ is the fermion number.  By
trading the Virasoro charges for mass and angular momentum
according to \bp, we see that the imaginary part of $\tau$ plays
the role of inverse temperature, while the real part is a chemical
potential for angular momentum.

Now consider $Z_{AdS}$.  We can again write a canonical formula
like \cd, but now its implementation is problematic since we lack
a satisfactory description of the Hilbert space in the
gravitational language.  At low energies the Hilbert space is well
understood as being comprised of a gas of particles moving on AdS,
but at sufficiently high energies we encounter black hole
solutions.  Black hole solutions are clearly not to be interpreted
as individual states of the theory (since they carry entropy), and
so it is not altogether clear how to include them in the trace.
The situation is more satisfactory in the path integral
formulation, where we can include the black holes as additional
saddle points of the functional integral, weighted by their
action.  Ultimately, since $Z_{CFT}$ {\it is} well defined, we
hope to use it to shed light on the Hilbert space of the
gravitational theory, including the black hole regime.

We will therefore attempt to make sense of
\eqn\ce{ Z_{AdS}(\tau,\taub) = \sum e^{-I}~.}
The summation is supposed to run over all saddle points of the
full effective action $I$ (which in principle includes all
corrections coming from string and loop corrections), such that
the boundary metric has modular parameter $\tau$. One subtle
point, whose importance will become clear as we proceed, is that
certain saddle points that are just coordinate transformations of
other saddle points will appear as distinct terms in the
summation. This is the analogue of the fact that in ordinary gauge
theories we can treat gauge transformations that are nontrivial on
the boundary  as global symmetries: we are not forced to demand
that physical states are invariant under such gauge
transformations.

Another important point is that we'll regard $I$ as capturing the
complete {\it local} part of the effective action.   This action
can in principle be computed in flat spacetime and then evaluated
on the asymptotically AdS backgrounds appearing in \ce.   But this
is not the whole story, as can be appreciated intuitively by
thinking in terms of field theory Feynman diagrams.   Geometries
contributing to \ce\ have a periodic imaginary time direction, and
there are contributions from Feynman diagrams that wind around the
time direction.  Such diagrams clearly are not incorporated in the
local effective action $I$.  Instead, we have to incorporate their
effects as additional saddle points in \ce.  In fact, there is a
clean way of isolating these effects via their behavior in $\tau$.
Local terms contribute to $\ln Z$ linearly in $\tau$ and $\taub$,
while the nonlocal terms are exponentially suppressed for ${\rm
Im}(\tau) \rightarrow 0$.  We will see this explicitly as we
proceed.

The simplest saddle point is just pure AdS$_3$ suitably
identified. We take the AdS$_3$ metric \bb\ and define  $w= \phi+
i t/\ell$, with $w$ identified as in \cc.   We know on account of
maximal symmetry that this is a saddle point of $I$, even taking
into account all higher derivative corrections.    What is the
value of  $I$ evaluated on this solution?  Since we don't know the
explicit form of $I$ we need to proceed indirectly.   The idea is
to integrate \bj.   To use \bj\ we need to work in coordinates
with fixed periodicity, so we define
\eqn\cf{ z= {i-\taub\over \tau -\taub}w -{i-\tau \over \tau
-\taub} \wb~,}
obeying $z\cong z+ 2\pi \cong z+2\pi i$.    $\tau$ now appears in
the metric,
\eqn\cg{ ds^2 = \left| {1-i\tau \over 2} dz +{1+i\tau \over 2}
d\overline{z} \right|^2~.}
Writing out \bj\ in the $z$ coordinates, and then converting back
to $w$ coordinates gives
\eqn\ch{ \delta I = 4\pi^2 i \left( - T_{ww} \delta \tau+
T_{\wb\wb}\delta \taub \right)~.}
Using that $L_0 = \Lt_0=0$ for this geometry, we know from \bn\
that
\eqn\ci{ T_{ww} = -{c \over 48\pi}~,\quad T_{\wb\wb} = -{\ct \over
48\pi}~.}
This yields the action
\eqn\cj{ I_{thermal} = {i \pi \over 12} (c\tau -\ct \taub)~.}

We can summarize the above computation as saying that we have
determined the exact low temperature behavior of $Z_{AdS}$.  In
particular, as ${\rm Im}( \tau) \rightarrow \infty$ we have
\eqn\ck{ \ln Z_{AdS}(\tau,\taub) =-{i \pi \over 12} (c\tau -\ct
\taub) + {\rm (exponentially~suppressed~terms)}~.}
This conclusion follows since we have incorporated all local terms
in the effective action, along with the fact that $L_0$ and
$\Lt_0$ have a gap in their spectrum above $0$ (this in turn
follows from the fact that AdS effectively acts like a finite size
box.) The exponentially suppressed terms are down by at least
$e^{2\pi i (\Delta \tau - \tilde{\Delta} \taub)}$, where $\Delta$
is the gap in the $L_0$ spectrum.  The contribution of these
suppressed  terms depends on the precise theory under
consideration (e.g. on the field content in addition to gravity)
and so we postpone incorporating them until later.

We'll now show that the high temperature behavior of the partition
function is governed by black holes.  To illustrate the basic
point in the simplest context, let's consider the non-rotating
black hole metric \bc.  In order to avoid a conical singularity at
$r=r_+$ we need to make the identification $t\cong t+ 2\pi
\ell^2/r_+$.    In other words, $\tau = i\ell/r_+$.  Note that we
have thereby  identified the Hawking temperature as $T = r_+/(2\pi
\ell^2)$.   So for $\tau$ purely imaginary the non-rotating black
hole metric contributes to the partition function as long as we
set $r_+ = i\ell/\tau$.

Next, we need to compute the action of the black hole to see if
and when it dominates the thermal AdS geometry.  As we'll now
show, after a judicious change of coordinates the needed
computation becomes equivalent to that yielding \cj.  We define
\eqn\cl{ w' = -{w\over \tau}~,\quad r' = {\ell \over r_+
}\sqrt{r^2 -r_+^2}~ }
with $w' = \phi'+it'/\ell$.    Then,  the black hole metric \bj\
becomes
\eqn\cm{ ds^2 =(1 + r'^2 /\ell^2 )dt'^2 + {\ell^2 dr'^2 \over 1+
r'^2 /\ell^2} +r'^2 d\phi'^2~,}
which is just the pure AdS$_3$ metric.  But now we have the
identifications $w' \cong w'+2\pi \cong w'+2\pi \tau'$, with
$\tau'=-1/\tau$.  In other words, we have shown the equivalence
(up to coordinate transformation) of thermal AdS with  modular
parameter  $\tau$ and a black hole with modular parameter
$\tau'=-1/\tau$:
\eqn\cn{ {\rm Thermal~AdS~with}~ \tau \quad \Leftrightarrow \quad
{\rm Black~hole~with}~\tau'=-1/\tau~.}
Of course, so far we have only established this for pure imaginary
$\tau$, but we'll generalize in due course.

Now, the action is invariant under the coordinate transformation
\cl, so we can immediately conclude that the black hole action is
\eqn\co{\eqalign{ I_{BTZ} = {i\pi \over 12}(c\tau' -\ct \taub')
=-{i\pi \over 12}({c\over\tau} -{\ct \over \taub} ) ~.}}
This results shows that at high temperature, ${\rm
Im}(\tau)\rightarrow 0^+$, the black hole has less action than
thermal AdS and hence will dominate the partition function.

\co\ gives the exact high temperature behavior of the partition
function; specifically, the part of $\ln Z$  linear in
$\tau^{-1}$,
\eqn\cp{ \ln Z ={i\pi \over 12}({c\over\tau} -{\ct \over \taub}) +
({\rm exponentially~suppressed ~terms)}~. }
Let's use this derive an expression for the entropy $S$ at high
temperature.   From \cd\ we can write in the saddle point
approximation
\eqn\cq{ \ln Z = S + 2\pi i\tau (L_0 -{c\over 24}) -2\pi i \taub
(\Lt_0  -{\ct \over 24})~.}
We further have
\eqn\crz{\eqalign{ L_0 - {c\over 24} &= {1 \over 2\pi i} {\p \ln Z
\over \p \tau} = -{c\over 24 \tau^2} \cr \Lt_0 - {\ct\over 24} &=
-{1 \over 2\pi i} {\p \ln Z \over \p \taub} = -{\ct\over 24
\taub^2}~. }}
From this we read off the entropy as
\eqn\cs{ S = 2\pi \sqrt{{c \over 6} (L_0 -{c \over 24})} +2\pi
\sqrt{{\ct \over 6} (\Lt_0 -{\ct \over 24})}~.}
This is the Cardy formula.  This formula in fact gives the high
temperature behavior of the entropy of any CFT, assuming unitarity
and a gapped spectrum of $L_0$ starting at $0$.  The standard
derivation of the Cardy formula is based on modular invariance,
and precisely parallels the gravitational approach adopted here;
so the agreement between the gravity and CFT sides is
unsurprising.  Indeed, this shows that the high temperature
entropy is guaranteed to agree between the two sides provided that
the central charges agree.

In the context of two-derivative gravity, formula \cs\ can be
written as $S=A/4G$.   But with higher derivative terms included
this is no longer the case.   More generally, we have Wald's
entropy  formula \wald\
\eqn\ctz{ S= -{1 \over 8 G} \int_{hor} \! dx \sqrt{h} {\delta
{\cal L} \over \delta
R_{\mu\nu\alpha\beta}}\epsilon^{\mu\nu}\epsilon^{\alpha\beta}~.}
In fact, \cs\ and \ctz\ are equivalent.\foot{Actually, \cs\ is a
bit more general in that \ctz\ only applies when $c=\tilde{c}$.}
This can be shown using \cb; for details, the reader is directed
to  \refs{\KrausVZ,\saidasoda}.

Consider the restricted case in which $c=\ct$.  Recall that the
value of $c$ is determined by extremizing the function \by. We can
translate this into an extremization principle for the entropy.
Specifically, $S$ is determined by extremizing the function \cs\
while holding fixed $L_0-{c\over 24}$ and $\Lt_0 -{\ct \over 24}$
(from \bp\ this is the same as holding fixed the dimensionless
mass and angular momentum).

Next, to set the stage for a more general discussion let's examine
the relation between thermal AdS$_3$  and the BTZ black hole from
a more geometrical perspective.   Thermal AdS$_3$ clearly has the
topology of a solid torus. The boundary is a two-dimensional
torus.  On the boundary torus there are two independent
noncontractible cycles, which we can take to be $\Delta \phi
=2\pi$, and $\Delta t/\ell  =  -2\pi i  \tau$ (we are just
considering the case of pure imaginary $\tau$ for the moment).
Now consider allowing these cycles to move off the boundary torus
into the bulk geometry.   It is then clear that the $\phi$ cycle
is contractible in the bulk while the $t$ cycle is not.

The coordinate transformation \cl\ that relates thermal AdS$_3$ to
BTZ interchanges $\phi$ and $t$, so for BTZ we find that it is the
$t$ cycle that  is contractible in the bulk, while the $\phi$
cycle is noncontractible.     This is illustrated in Fig. 1
\fig{Relation between BTZ and thermal
AdS$_3$}{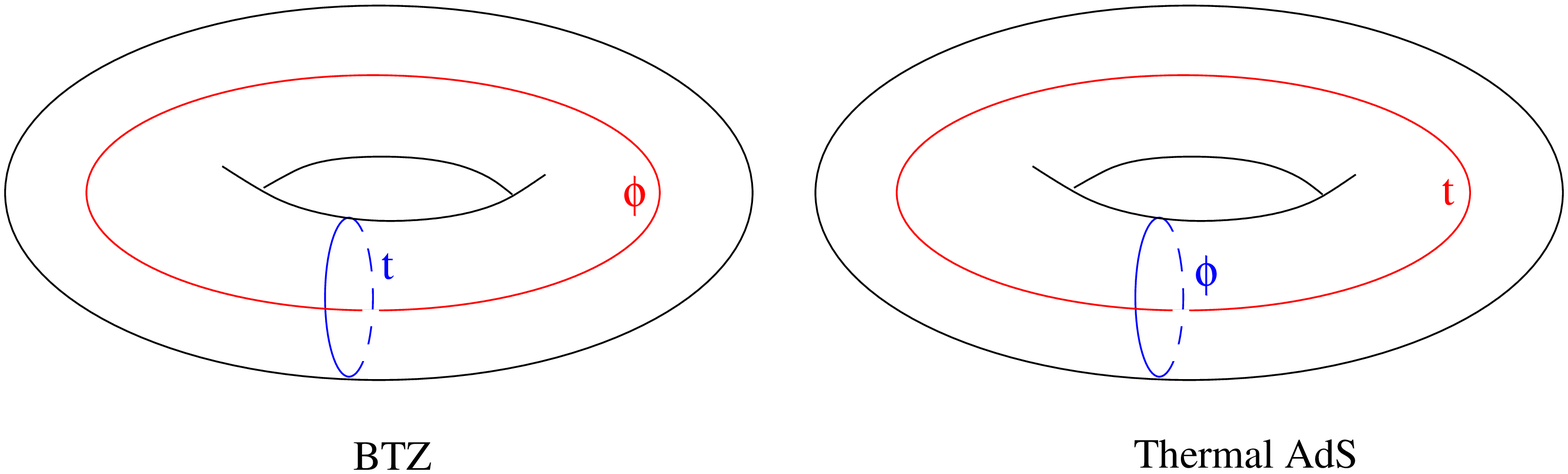}{7.0truein}

The generalization of this story involves rotating black holes.  To deal
with this efficiently it is advantageous to describe the BTZ black hole as
a quotient of AdS.

\subsec{BTZ black holes as quotients \CarlipGC}

Euclidean AdS$_3$ can be written
\eqn\cu{ds^2 = {d\rho^2+ dz d\zb \over \rho^2}~.}
Now consider the matrix
\eqn\cv{ g = \left(\matrix{ \rho+{z \zb / \rho} & {z / \rho} \cr {\zb / \rho}
& {1 / \rho} } \right)~.}
$\det g =1$, so $g \in SL(2,\IC)$.   Actually, $g$ can be written
as $g =hh^\dagger$ with $h\in SL(2,\IC)$.   Since $g$ is invariant
under  $h\rightarrow hf$ with $f \in SU(2)$, we see that the space
of $g$ matrices can be identified with the coset
$SL(2,\IC)/SU(2)$.   The line element \cu\ is the same as the
natural line element on the coset,
\eqn\cw{ ds^2 = \half \Tr  (g^{-1} dg g^{-1} dg)~.}
Since this is invariant under $h\rightarrow \alpha h$,  $\alpha
\in SL(2,\IC)$, we see that Euclidean AdS$_3$ has an $SL(2,\IC)$
group of isometries.

The BTZ black hole is obtained from AdS$_3$ by making  $SL(2,\IC)$
identifications, $h \cong \gamma h$.     Since we can always
redefine $h$ as $h=\alpha h'$, we see that $\gamma$ is only
defined up to conjugation by $SL(2,\IC)$.   So without loss of
generality we can take $\gamma$ to be diagonal and write
\eqn\cx{ \gamma = \left( \matrix{ e^{-i \pi \tau} & 0  \cr 0&
e^{i\pi \tau} }\right)~.}
In terms of coordinates, this implies the identification
$(\rho,z)\cong (e^{-i\pi (\tau-\taub)}\rho, e^{-2\pi i \tau} z)$.
Now write $z=e^{-iw}$ so that at the boundary ($\rho=0$) we have
the identification $w\cong w + 2\pi \tau$.     This identifies
$\tau$ in \cx\ as the modular parameter of the boundary torus.
In other words, quotienting AdS$_3$ by $\gamma$ yields thermal
AdS$_3$:
\eqn\cy{ ds^2 = (1+r^2/\ell^2)dt^2 +{dr^2 \over 1+r^2/\ell^2} + r^2 d\phi^2 }
with the identification $w\cong w+2\pi \cong w+2\pi \tau$, where
$w= \phi+it/\ell$.

To construct an ``$SL(2,\IZ)$ family of black holes"
\refs{\MaldacenaBW,\farey}  we consider the modular transformed
version of \cx
\eqn\cz{ \gamma = \left( \matrix{ e^{-i \pi {a\tau+b \over c\tau+d}} & 0 \cr 0& e^{i\pi{a \tau+b \over c\tau+d}} }\right)~,}
with $(a,b,c,d)\in \IZ$ and  $ad-bc=1$.  This is   a geometry
whose conformal boundary has modular parameter ${a\tau+b \over
c\tau+d}.$  But, as we illustrated in a simplified context above,
if we change coordinates we can bring the modular parameter back
to $\tau$. The action for this geometry can be read off from \cj,
\eqn\da{ I(\tau,\taub) = {i \pi \over 12} \left[ c  \left({a\tau
+b \over c\tau +d}\right) - \tilde{c}\left({a\taub +b \over
c\taub+d}\right)\right]~.}
Note the unfortunate notation in which the same symbol $c$ appears
with two different meanings.  These $SL(2,\IZ)$  black holes will
make an appearance later as saddle points of the Euclidean path
integral.

For completeness, we will explicitly write the metric  of the
rotating BTZ black hole, generalizing \bc. Start from \cu\ and
write
\eqn\db{\eqalign{ z& = \left({r^2-r_+^2 \over r^2 -r_-^2}
\right)^{1/2}\exp \left\{ {r_+ +r_- \over \ell }(\phi+i
t/\ell)\right\} \cr \rho & = \left({r_+^2-r_-^2 \over r^2
-r_-^2}\right)^{1/2}\exp \left\{ {r_+ \over \ell}\phi +ir_- t/\ell
\right\}  }}
with $r_+$ real  and $r_-$ imaginary.   The metric is then

\eqn\dc{ ds^2 = {(r^2 -r_+^2)(r^2 -r_-^2) \over r^2
\ell^2}dt^2+{\ell^2 r^2 \over
(r^2-r_+^2)(r^2-r_-^2)}dr^2+r^2(d\phi + i{r_+  r_- \over \ell r^2}
dt)^2~.}
The modular parameter is
\eqn\dd{ \tau = {i \ell \over r_+ +r_-}~.}
The action is given by \co.  The event horizon is at  $r=r_+$  and
the entropy is given by \cs\ with
\eqn\de{ L_0-{c\over 24} = {(r_+-r_-)^2 \over 16 G \ell}~, \quad
\Lt_0-{\ct\over 24} = {(r_++r_-)^2 \over 16 G \ell}~.}
To get the Lorentzian black hole we replace $t$ by $it$ and
continue $r_-$ to real values.  Note that the Virasoro charges
\dc\ are then real.   Extremal black holes correspond to $r_- =
\pm r_+$.

\newsec{Charged black holes and Chern-Simons terms}

An important generalization is to allow our black holes to carry
charge.  We'll see that this comes about in an elegant fashion. To
begin, we'll consider a collection of $U(1)$ gauge fields. Besides
the usual Maxwell term, since we are working in an odd-dimensional
spacetime our gauge fields can have Chern-Simons terms. We'll find
that the Chern-Simons terms  are needed in order for the charge to
be nonzero. Indeed, the charge comes {\it entirely} from the
Chern-Simons terms.  Later, we'll indicate where these
Chern-Simons originate from the higher  dimensional
string/M-theory viewpoint.  Our presentation  follows \KrausNB.
Additional relevant work on Chern-Simons theory includes
\refs{\ElitzurNR,\GukovID}.

\subsec{$U(1)$ gauge fields in AdS$_3$}

Let's first consider the case of a single $U(1)$ gauge field.   A
Chern-Simons term in three spacetime dimensions is, in
differential form language,
\eqn\df{I_{CS} = {i k \over 8\pi} \int_M \! AdA~.}
The normalization was chosen   so that the constant $k$ will be
identified with the level of a  corresponding  current algebra. A
gauge transformation is $\delta A = d\Lambda$.    $I_{CS}$ is not
gauge invariant, but instead  varies by a boundary term:\foot{Note
 that the equations of motion will still be gauge invariant.}
$\delta I_{CS} = {i k \over 4\pi}\int_{\p M} \Lambda dA$.

 The gauge field admits  an expansion analogous to \bh,
\eqn\dg{ A = A^{(0)} + e^{-2\eta/\ell} A^{(2)} +\ldots~,}
and we choose the gauge $A_\eta =0$.    Analysis of the field
equations (including the effect of Maxwell type terms) shows  that
$A^{(0)}$ is a flat connection; that is, the field strength
corresponding to \dg\ falls off as $e^{-2\eta/\ell}$.  For this
reason, all the results we derive below for the currents and
stress tensor will be valid in the presence of arbitrary higher
derivative terms for the gauge fields.  The flatness of $A^{(0)}$
implies that only the Chern-Simons terms yields nonzero boundary
variations.

In analogy with \bj\  a boundary current is obtained from the
on-shell variation of the action with  respect to $A^{(0)}$,
\eqn\dh{ \delta I = {i\over 2\pi}  \int_{\p AdS}\! d^2x
\sqrt{g^{(0)}}J^{\alpha }  \delta A^{(0)}_\alpha ~.}
We now need to define the appropriate variational principle that
yields the equations of motion for our gauge field.  For
definiteness, consider the pure AdS geometry \bb. Naively, one
might guess that in the variational principle one could hold fixed
both $\A0_t$ and $\A0_{\phi}$. But this is too strong, since there
will then typically be no smooth solutions of the equations of
motion with the assumed boundary conditions.   The issue is the
holonomy around the contractible $\phi$ circle, expressed by $\int
\! d\phi A_\phi$. When we contract the circle we need the holonomy
to either vanish or  match onto  an appropriate source to avoid a
singularity. So it is only $\A0_t$ that can take generic values.
If we define $w=\phi +it/\ell$ as usual, then an appropriate
variational principle is to hold fixed {\it either} $\A0_w$ or
$\A0_{\wb}$, but not both.      The sign of $k$ will determine
which component to hold fixed.

Let's assume that $k$ is positive.  Then we claim that to derive
the equations of motion  we should demand that the action be
stationary under variations that hold fixed  $\A0_{\wb}$.  That is
to say, we demand that \dh\ take the form
\eqn\di{ \delta I = {i\over
2\pi}  \int_{\p AdS}\! d^2w \sqrt{g^{(0)}}J^{\wb } \delta A^{(0)}_{\wb} ~.}
But we can readily check that the variation of \df\ does not take this form.
However, we still have the freedom to add boundary terms to the action.
If we add to the action the term
\eqn\dj{ I^{bndy}_{gauge} = -{k \over 16\pi} \int_{\p AdS}\! d^2x \sqrt{g}
g^{\alpha \beta} A_\alpha A_\beta }
then the variation of the action {\it does} take the form \di\ with
\eqn\dk{ J_w = \half J^{\wb} = {ik\over 2} \A0_w~.}
Another way to say this is that $J_{\wb}=0$, which means that our current
is purely leftmoving.

Note that the boundary term \dj\ depends on the metric, and so
will contribute to the stress tensor.   This in contrast to the
topological term \df.   A straightforward computation yields
\eqn\dl{ T^{gauge}_{\alpha\beta}  =  {k \over 8\pi}(A^{(0)}_\alpha
A^{(0)}_\beta
 -\half A^{(0)
\gamma}A^{(0)}_\gamma g^{(0)}_{\alpha\beta})~,}
or, in complex coordinates,
\eqn\dm{\eqalign{ T^{gauge}_{ww} &=  {k \over 8\pi} A^{(0)}_w
A^{(0)}_{w}~, \cr T^{gauge}_{\wb \wb} &= {k \over 8\pi}
A^{(0)}_{\wb} A^{(0)}_{\wb} ~, \cr
 T^{gauge}_{w \wb}&=T^{gauge}_{ \wb w}=0~.}}
The index (0) on the gauge field reminds us that  boundary
expressions strictly refer to just the leading term in the
expansion \dg\ for the bulk gauge field. In the following we will
reduce clutter by dropping this index.

 We can now
see why the sign of $k$ is important; if we took  $k$ negative
\dm\ would imply that the energy is unbounded below.   The case of
$k$ negative needs to handled differently, by flipping the sign of
the boundary term   \dj.    The same analysis then yields a purely
rightmoving current.

Turning to the general case of multiple $U(1)$ gauge fields, we
write the action as
\eqn\dn{I=  {i \over 8\pi} \int \! d^3x \, (k^{IJ}  A_I
dA_J-\kt^{IJ} \At_I d\At_J) - {1 \over 16\pi} \int_{\p AdS} \! d^2
x \sqrt{g} g^{\alpha\beta} (k^{IJ}A_{I \alpha}
A_{J\beta}+\kt^{IJ}\At_{I\alpha} \At_{J\beta} )   ~.}
Both $k^{IJ}$ and $\kt^{IJ}$ are symmetric matrices with positive
eigenvalues. The $IJ$ indices on   $k^{IJ}$ versus $\kt^{IJ}$ are
independent,  and so can take different ranges.

In conformal gauge, the gauge fields contribute to the currents
and stress tensor as,
\eqn\do{\eqalign{ T^{gauge}_{ww} &= {1 \over 8\pi} k^{IJ} A_{Iw} A_{Jw}+
 {1 \over 8\pi} \kt^{IJ} \At_{Iw}\At_{Jw} ~, \cr
T^{gauge}_{\wb \wb} &=  {1 \over 8\pi} k^{IJ}A_{I\wb}A_{J\wb}+
{1 \over 8\pi}\kt^{IJ} \At_{I\wb}\At_{J\wb}~,
\cr
 T^{gauge}_{w \wb}&=T^{gauge}_{ \wb w}=0~, \cr
 J^{I}_{ w} &= {i\over 2}k^{IJ} A_{Jw}~,\quad
 J^{I}_{\wb}=0~, \cr \Jt^{I}_{ w}&=0~,\quad  \Jt^{I}_{ \wb} = {i\over 2}\kt^{IJ} \At_{J \wb}
 ~.}}

The modes of the currents are defined as
\eqn\dpp{ J^I_n = \oint \!{dw \over 2\pi i} e^{-inw} J^I_w~,\quad
 \Jt^I_n = -\oint \!{d\wb \over 2\pi i} e^{in\wb} \Jt^I_{\wb}~.}
By writing out the formulas for the changes in the  stress tensor
and currents under a variation of the gauge field we can infer the
commutation relations
\eqn\dq{\eqalign{ [L_m,J^I_n] & =-n J^I_{m+n} \cr
[J^I_m,J^J_n]&= \half m k^{IJ} \delta_{m+n}~,}}
and likewise for the tilded generators.

\subsec{Spectral flow}

Together with the Virasoro algebra \bo\ the algebra \dq\ admits  a
so-called ``spectral flow" automorphism that will play an
important role.  For arbitrary parameters $\eta_I$ the algebra is
preserved under
\eqn\dra{\eqalign{ L_n &\rightarrow L_n + 2\eta_I J^I_n + k^{IJ} \eta_I \eta_J \delta_{n,0} \cr
J^I_n &\rightarrow J^I_n +k^{IJ} \eta_J \delta_{n,0}~.}}
From our explicit formulas for the generators we see that this  is
equivalent to
\eqn\ds{ A_{Iw} \rightarrow A_{Iw} + 2\eta_I~.}
This constant shift of the gauge potentials is equivalent to
shifting the periodicities of charged fields.  In particular,
since  the phase factor acquired by a particle of charge $J^I_0$
taken around the AdS cylinder is
\eqn\dt{ e^{i J^I_o \oint \! dw A_{Iw} }~,}
we see that the shift \ds\ induces the the phase $e^{4\pi i J^I_i \eta_I}$.

From now on we will use another normalization for the gauge charges by
defining
\eqn\du{ q^I =2J^I_0~,\quad \qt^I =2 \Jt^I_0~,}
where the $2$ is introduced for convenience.

\subsec{Nonabelian gauge fields}

Besides the $U(1)$ gauge fields, in the main cases of interest
from the string theory perspective we will also have $SU(2)$ gauge
fields. From the higher dimensional point of view we will be
considering either AdS$_3\times S^2$ or AdS$_3 \times S^3$
geometries.  The spheres have isometry group $SO(3) \cong SU(2)_R$
or $SO(4)\cong SU(2)_L \times SU(2)_R$, and we then have the
associated Kaluza-Klein gauge fields.  To show that these gauge
fields have three-dimensional Chern-Simons terms is somewhat
subtle, but can be derived by a careful consideration of the
background flux configuration that supports the sphere
\refs{\HMM,\HansenWU}.

The $SU(2)_L$ Chern-Simons term is
\eqn\dv{ I_{CS} =-{ik\over 4\pi} \int \! d^3x \,\Tr(AdA+{2\over 3} A^3)~,}
with $A = A^a {i\sigma^a \over 2}$.   Invariance of  the path
integral under large gauge transformation fixes $k$ to be an
integer.    The $SU(2)_R$ Chern-Simons term is taken with the
opposite sign, as above.   As before, in order to get purely left
or right moving currents we need to add a boundary tern.   This
has the same form as \dj\ except that we sum over the group
indices. We will just be considering solutions in which $A^{(0)a}$
and $\At^{(0)a}$ are nonvanishing only for $a=3$.     We can then
easily incorporate the corresponding currents into the previous
discussion by extending the $I$ index to include $I=0$, and write
\eqn\dw{ A^{(0)3}=A_{I=0}~,\quad k=k^{00}~,\quad k^{0,I>0}= k^{I>0,0}=0~,}
and likewise for the tilded counterparts.   All the formulas \do\
now carry over.

\subsec{Supersymmetry}

We now discuss one important  implication of supersymmetry.   In
two dimensions we  characterize the amount of supersymmetry by the
number of left and right moving supercharges, $(N_L,N_R)$.   Here
the focus will be on theories with either $(0,4)$ or $(4,4)$
supersymmetry. Supersymmetry then implies the existence of an
$SU(2)$ R-symmetry that rotates the supercharges into one another
(the $4$ supercharges transform as two doublets).   In the $(4,4)$
case we have $SU(2)_L \times SU(2)_R$ R-symmetry.   The R-symmetry
currents correspond to the $SU(2)$ gauge fields in \dv.

Of central importance to us is that the supersymmetry algebra
relates the level of the $SU(2)$ current algebra $k$ to the
central charge $c$ as $c=6k$.    When we recall that $k$ appears
in \dv, we see that determining the exact central charge is
equivalent to determining the $SU(2)$ Chern-Simons term.    Of
course, in the $(0,4)$ case this argument only gives us the right
moving central charge, but we will see momentarily how a related
argument gives the left moving central charge.

\subsec{Gravitational Chern-Simons term \KrausZM}

The left and right moving central charges of a two-dimensional CFT
are independent and need not be equal.    However, if $c\neq \ct$
it is not possible to couple such a theory to gravity in a
diffeomorphism invariant fashion: there  is a gravitational
anomaly \refs{\AlvarezGaumeIG,\BardeenPM,\GinspargQN}.  To write
the anomaly we can work in terms of the connection 1-forms,
defined as $\Gamma^i_{~j} = \Gamma^i_{jk} dx^k$, where
$\Gamma^i_{jk}$ are the usual Christoffel symbols. The breakdown
of diffeomorphism invariance is signaled by the non-conservation
of the stress tensor:\foot{Alternatively, one can add a local
counterterm to render the stress conserved but non-symmetric.}
\eqn\dx{ \nabla_i T^{ij} =-i {c - \ct \over 96 \pi} ~g^{ij}
\epsilon^{kl} \p_k \p_m \Gamma^m_{il}~.}
Equivalently, under an infinitesimal diffeomorphism, $x^i
\rightarrow x'^i= x^i - \xi^i(x)$, the effective  is not
invariant, but instead changes by
\eqn\dy{ \delta I_{eff} = -i{c - \ct \over 96 \pi} ~\int \Tr  (v
d\Gamma)~,}
with $v^{i}_{~j} = \p_j \xi^i$.

To reproduce this from the AdS point of view we need a term in the
bulk that varies under diffeomorphisms.  Further, the variation
should be a pure boundary term, otherwise the bulk theory will be
inconsistent.  Up to boundary terms, there is a unique
possibility, the gravitational Chern-Simons term \DeserWH:
\eqn\dz{ I_{CS}(\Gamma) =-i\beta \int\! \Tr (\Gamma d\Gamma  +{2
\over 3}\Gamma^3)~.}
Under an infinitesimal diffeomorphism $\delta \Gamma = dv  +
[\Gamma,v]$, so
\eqn\ea{ \delta I_{CS} =-i \beta \int_{\p AdS} \Tr (vd\Gamma)~.}
Comparing with \dy\ we read off
\eqn\eaa{\beta = {c-\ct \over 96\pi}~.}
The full stress tensor in the presence of the gravitational
Chern-Simons term is discussed in \refs{\KrausZM,\SolodukhinAH}.

We now know how to compute both central charges of the $(0,4)$
theory. The coefficient of the $SU(2)$ Chern-Simons term gives us
$\ct$, while the gravitational Chern-Simons term gives us $c-\ct$.
The reason why this is useful is that (at least in the cases we'll
consider) the Chern-Simons terms arise at tree level and one-loop,
but receive no other corrections, and hence we can determine them
exactly. In particular, we can determine the central charges
without knowledge of the full effective action including higher
derivative terms.  This is to be contrasted with the
non-supersymmetric case, where the c-extremization procedure
discussed in section 2.3, while efficient, does  require the
explicit action as an input.

\newsec{String theory constructions}

We will now review two standard constructions of AdS$_3$
geometries in string theory.   The first,  and best known, example
is realized as the near horizon geometry of the D1-D5 system
\StromingerSH\ (for reviews see \refs{\AharonyTI,\DavidWN}). This
yields AdS$_3 \times S^3 \times M_4$, where $M_4$ is $T^4$ or
$K3$.   The second example \MaldacenaDE\ is realized in terms of
wrapped M5-branes, and yields AdS$_3 \times S^2 \times M_6$, with
$M_6$ being $T^6$, $K3\times T^2$, or $CY_3$.

\subsec{D1-D5 system}

To describe the brane construction, we first work at weak string
coupling and consider  IIB string theory on $R^{4,1} \times S^1
\times M_4$.  We wrap $N_5$ D5-branes on $S^1 \times M_4$, and
$N_1$ D1-brane on $S^1$.  This setup preserves $8$ of the original
$32$ supercharges.   When the length scale associated with  $M_4$
is small compared to the $S^1$ the low energy dynamics of the
system is described by a theory  on the $1+1$ dimensional
intersection.   The standard weak coupling open string
quantization yields a  $U(N_1) \times U(N_5)$ supersymmetric gauge
theory, which flows to a nontrivial CFT in the infrared, with
$(4,4)$ susy.   We want to know the left and right moving central
charges.   The easiest way to compute these is by using anomalies
(see section 5.3.1 of \AharonyTI). Focus on the left moving side,
say. As we have discussed previously, the existence of $4$
supercharges means that the CFT has an $SU(2)$ R-symmetry.  The
level $k$ of the corresponding current algebra is related to the
central charge by $c=6k$. The current algebra also implies that
the R-symmetry currents are anomalous when coupled to external
gauge fields,\foot{There is some choice  in the form of the right
hand side due to the freedom to add non-gauge invariant  local
counterterms to the action. }
\eqn\ee{D_{\wb}J^a_w = {ik\over 2} \p_w A_{\wb}^a  ~.}
    Chiral anomalies
are related to topology, and are invariant under smooth
deformations of the theory. In our context, the level $k$ is an
integer, which can equally well be evaluated in the weak coupling
gauge theory description valid in the UV.   The anomaly arises
from one loop diagrams.  This is a fairly straightforward
computation, and the result is that $k= N_1 N_5$.  The same
analysis holds for the right movers.  We conclude that the exact
central charges are $c=\ct =6N_1 N_5$.    Note how little went
into this result: we just needed to know that  in the IR we have
$(4,4)$ susy, and that the IR theory is reached via RG flow from
the UV gauge theory.

Next, we recall the anomaly inflow mechanism \CallanSA, which will
be useful in relating the CFT central charges to those of the
AdS$_3$ theory.    Note that the D1-D5 system is localized at a
point in the $4$ noncompact spatial dimensions, and hence is
invariant under the  corresponding $SO(4)$ group of rotations.  If
we write $SO(4) \cong SU(2)_L \times SU(2)_R$, we identify the
left and right moving $SU(2)$ R-symmetry groups.       We can
think of the $SO(4)$ as acting on the vector space normal to the
brane worldvolume (the so-called ``normal bundle"). We can further
allow the $SO(4)$ rotations to vary over the worldvolume, which
leads to an $SO(4)$ gauge theory.   Now, we know from \ee\ that in
general the worldvolume theory on the brane is not invariant under
such local $SO(4)$ transformations, the effective action instead
varies as
\eqn\ef{ \delta I_{brane} = -{i\over 4\pi} \int\! d^2 w  \left(
D_{\wb} J^a_w \Lambda^a + D_w \Jt^a_{\wb} \tilde{\Lambda}^a
\right)~,}
where we have written the result in terms of the $SU(2)_L \times SU(2)_R$
parameters.

 On the other hand,  from the point of view of the full ten
dimensional string theory these gauge transformations are just
coordinate transformations (or, more accurately, local Lorentz
transformations), but it is well known   that the full theory is
nonanomalous.   Indeed, otherwise the IIB string theory would be
inconsistent since we are talking about a potential breakdown of
diffeomorphism invariance.  So something must be cancelling the
variation of the brane effective action.   Now, the entire theory
consists of the theory on the branes coupled to the bulk ten
dimensional fields.  We therefore conclude that the bulk theory
must have an $SO(4)$ variation that cancels that of the brane
theory. The details of this have been worked out  in various
examples \HMM\ (although not explicitly for the D1-D5 system, to
our knowledge), and we will examine this in more detail in our
next example.   One finds that there is an inflow of current from
the bulk region onto the branes that reproduces the anomalous
divergence of the brane current.    Here we will just take for
granted that such a cancellation indeed takes place for the D1-D5
system coupled to the IIB theory.

We now turn to the supergravity description of the D1-D5 system.
The  starting point is the action of IIB supergravity.  We will
only write down the dependence on the metric, dilaton, and RR
3-form field strength, since the other fields will be vanishing in
the solution.  In particular, this means  that we need not concern
ourselves with the subtleties associated with the self-duality of
the 5-form field strength.
\eqn\eb{I = {1 \over 2 \kappa^2_{10}}\int\! d^{10}x \sqrt{G} e^{-2\Phi}\left(R + 4 (\p \Phi)^2
 +\half e^{2\Phi}  |G_3|^2 \right)~.}
The action is written  in Euclidean signature and in  terms of the
string frame metric.

The equations of motion admit the following solution  representing
the D1-D5 system,
\eqn\ec{\eqalign{ds^2 & =(Z_1 Z_5)^{-1/2}(dt^2 + dx_5^2)  + (Z_1
Z_5)^{1/2} dx^i dx^i + (Z_1/Z_5)^{1/2}      ds_{M_4}^2 \cr G_3 &=2
Q_5 \epsilon_3 + 2 iQ_1 e^{-2\Phi} \star_6 \epsilon_3 \cr
e^{-2\Phi} & = Z_5/Z_1~.}}
Here $\epsilon_3$ is the volume form on the unit 3-sphere, and
$\star_6$ is the Hodge dual in six dimensions.  The branes
intersect  over $(t,x_5)$, and we denote the four noncompact
spatial direction by $x^i$.
 The harmonic functions $Z_{1,5}$ are
\eqn\ed{ Z_{1,5} = 1+ {Q_{1,5} \over r^2}~,\quad Q_1 =  {(2\pi)^4
g N_1 \ap^3 \over V_4}~, \quad Q_5 =  g N_5 \ap~.}
To isolate the near horizon geometry we drop the $1$ from  the
harmonic functions, and arrive at
\eqn\ee{\eqalign{ds^2 &= {r^2 \over \ell^2}(dt^2 +dx_5^2)+{\ell^2
\over r^2}dr^2 + \ell^2 d\Omega_3^2 +(Q_1 /Q_5)^{1/2}
ds_{M_4}^2\cr G_3 & = 2 Q_5(\epsilon_3  +i \star_6 \epsilon_3) \cr
e^{-2\Phi} &= Q_5/Q_1  ~.}}
with
\eqn\ef{\ell^2 =( Q_1 Q_5)^{1/2}~.}

A change of coordinates brings the $(t,x_5,r)$ part of the metric
to the form \bb,\foot{Actually, since in our case $x_5$ is compact
we only get \bb\ locally. To get  precisely \bb\ one should
instead start  with a rotating version of the D1-D5 metric
\refs{\CveticXH, \BalasubramanianRT,\MaldacenaDR}.} and so we
recognize the geometry as AdS$_3 \times S^3 \times M_4$.

According to the Brown-Henneaux formula \bl\  the central  charge
is given by $c={3 \ell \over 2 G }.$  Here $G$ refers to the three
dimensional Newton's constant, which we compute as
\eqn\eg{ G = G_{10} {1 \over V_7} = 8\pi^6 \ap^4 g^2 e^{2\Phi} {1 \over 2\pi^2 \ell^3 (Q_1/Q_5) V_4}~.}
Therefore
\eqn\eh{ c =6 N_1 N_5~,  }
in agreement with the microscopic result.

We now ask {\it why} the microscopic and supergravity
computations of the central charges agree.  Underlying the
agreement is the conjectured AdS/CFT duality between the two
descriptions, but it is more satisfying to give a direct argument
using just known facts. The key to this will be the relation
between $c$ and $k$, as well as the anomaly inflow mechanism.
Compare the weak coupling description of our system, as a bound
state of branes sitting in an ambient flat spacetime, to the
supergravity description, as   a near horizon AdS geometry joined
to an asymptotically flat region. We can think of interpolating
between these two pictures by dialing the string coupling.  In the
brane description we convinced ourselves that the chiral anomaly
of the brane theory was cancelled by the bulk theory via an inflow
of current.     Now consider what happens as we increase the
string coupling.  The brane anomaly is fixed in terms of the
integer $k$, and hence does not change, implying also that the
bulk inflow is unchanged as we increase the coupling.  But in the
supergravity description we join this same asymptotic bulk
geometry onto the near horizon region.   In the supergravity
picture we separate the two regions by an artificial border.  It
is clear that the current flows smoothly across the border, and
there are no sources or sinks of currents there.   We conclude
that the current inflow from the asymptotic region must precisely
equal that into the near horizon region.   But now, putting things
together, we have found that the anomaly of the brane theory must
precisely match the AdS current inflow.   Since both are
determined by a parameter $k$, it must be the case that $k_{CFT} =
k_{sugra}$, which is what we were trying to establish. To
summarize, the matching of $k$, and hence $c$, between the CFT and
gravity descriptions is dictated by anomaly cancellation.   A
mismatch between the two would imply that  IIB string theory is
anomalous, which we know not to be the case.

This does not yet explain why  our gravity computation yielding
\eh\ agrees {\it exactly} with the CFT result, since we started
from the two-derivative approximation to the supergravity action.
Our anomaly argument really applies to the full action, with all
higher derivative terms included. But in general, instead of using
$c=3\ell/2G$, we should instead use \cb, which in general receives
correction from higher derivative terms.   However, the spirit of
the anomaly argument suggests that we are better off computing
$k$.  We will now outline this computation.

To compute $k$ we need to find the  coefficient of the
Chern-Simons term \dv. The importance of this term is that under a
gauge transformation, $\delta A = d\Lambda +[\Lambda, A]$,  we
have
\eqn\ei{\delta I_{CS} = -{ik\over 4\pi} \int_{\p AdS}  \! \Tr
(\Lambda dA)~.}
This is the result for one of the $SU(2)$ factors of  the $SO(4)$
gauge group; the other factor gives the same result except with a
flipped sign. Our strategy is then to similarly compute the gauge
variation of \eb\ and compare this to \ei\ to read off $k$.

This turns out to be more subtle than one might have expected.  We
will sketch the main points, directing the reader to \HansenWU\
for a more thorough treatment. The first step is to identify the
$SO(4)$ gauge fields, which we recall are the Kaluza-Klein gauge
fields associated with rotations of the $S^3$.   Instead of the
pure AdS$_3 \times S^3 \times M_4$ metric in \ee\ we make the
replacement
\eqn\ej{ d\Omega_3^2 ~\rightarrow ~ (dy^i -A^{ij}y^j)(dy^i-A^{ik}y^k)~,}
with $y^{1,2,3,4}$ obeying $\sum_{i=1}^4 y^i y^i =1$.   The
1-forms $A^{ij} =-A^{ji}$ are the KK gauge fields, and are allowed
to have dependence on the AdS coordinates $(t,x_5,r)$.    To be
more precise, we just impose this form of the metric
asymptotically as we approach the AdS boundary.    Note that the
metric is invariant under
\eqn\ek{ A^{ij} \rightarrow d\Lambda^{ij} +[\Lambda, A]^{ij}~,
\quad y^i \rightarrow y^i + \Lambda^{ij} y^j~.}

Given this asymptotic form of the  metric we also need to specify
the asymptotic form of $G_3$, generalizing that in \ee.  This is
where the subtlety lies. We need $G_3$ to be closed, and to have a
fixed integral over $S^3$, since this is the D5-brane charge.   We
might also try to have $G_3$ invariant under \ek. If so, we would
then find that the action is gauge invariant (i.e. invariant under
$\delta A =d\Lambda +[\Lambda, A]$), as follows from the
diffeomorphism invariance of \eb\ combined with the invariance
under \ek.   It turns out that it is not possible to satisfy all
these conditions simultaneously, and the best one can do is to
find a $G_3$ that varies under \ek\ as
\eqn\el{\delta G_3  = {1 \over 4} Q_5 \Tr( d\Lambda dA -d\tilde{\Lambda} d\tilde{A})~.}
While this implies that the action \eb\ is not gauge invariant,
it can be shown that
 the variation is a
boundary term, so the equations of motion are gauge invariant.
In particular, the explicit computation yields
\eqn\em{\delta I  = -{ik \over 4\pi} \int_{\p AdS} \! \Tr (
\Lambda dA -\tilde{\Lambda} d\tilde{A})~,}
with $k=N_1 N_5$.   This then reproduces $c=6k = 6N_1 N_5$.

To complete this circle of ideas we  should now show that this
conclusion is unchanged under the addition of higher derivative
terms to \eb.   This has not yet been demonstrated explicitly.
Here we will just note that the condition that the action vary
only by a boundary term greatly restricts the form of the action,
and makes it plausible that there are no further corrections.  In
any case, in the example discussed in the next section involving
M5-branes we will give the  complete derivation.

\subsec{Wrapped M5-branes}

Our other example of an AdS$_3$ geometry arises from wrapping
M5-branes on a 4-cycle in $M_6$, where $M_6$ can be $T^6$,
$K3\times T^2$ or $CY_3$. Starting from the eleven dimensional
M-theory compactified on $M_6$, this produces a string like object
in the five noncompact directions.  We further  compactify the
direction along the string, to leave four noncompact directions.
This system was studied extensively in
\refs{\MaldacenaDE,\MinasianQN}.

Letting  $\Omega_I$ be a basis of 4-cycles in $M_6$, we take the
M5-brane to wrap $P = p^I \Omega_I$.     At low energies the
theory on the resulting string   flows to a  CFT with $(0,4)$
susy.      The left and right moving central charges can be
computed by studying the massless fluctuations of the M5-brane,
including the self-dual worldvolume 3-form field strength.   The
result of this analysis is
\eqn\en{ c = C_{IJK}p^I p^J p^K + c_{2I} p^I~,\quad \ct
=C_{IJK}p^I p^J p^K + \half c_{2I} p^I~.}
$C_{IJK}$ denotes the number of triple intersections of the three
4-cycles labelled by $I,J,K$ (note that three 4-cycles generically
intersect over a point in six dimensions.)    $c_2$ is the second
Chern class of $M_6$ which we can expand in a basis of 4-forms
with expansion coefficients $c_{2I}$.    For our purposes, the
main point is that $C_{IJK}$ and $c_{2I}$ are certain topological
invariants, and so we see that the central charges are moduli
independent.

Instead of studying the fluctuation problem, we can compute the
central charges from the anomaly inflow mechanism \HMM.   The
relevant anomalies are those with respect to the right moving $
SU(2)$ R-symmetry, and with respect to worldvolume
diffeomorphisms. The relevant terms in the eleven dimensional
action are
\eqn\eo{\eqalign{ 2\kappa_{11}^2 I &= \int\! d^{11}x
\sqrt{g}(R+\half |F_4|^2) +{i \over 6} \int\! A_3 \wedge F_4
\wedge F_4 \cr & \quad +{i(2\kappa^2_{11})^{2/3} \over 3 \cdot
2^{6} \cdot (2\pi)^{10/3} }\int\! A_3  \wedge \left[\Tr R^4 - {1
\over 4} (\Tr R^2)^2\right]~.}}
The terms in the first line are the standard two derivative
bosonic terms.  In the second line we have written a particular
eight derivative term. Of course, there are an infinite series of
other higher derivative terms (see
\refs{\generalRRRR,\tseytRRRR,\antRRRR} for some general results),
but the important point is that in \eo\ we have written the only
two Chern-Simons terms (i.e. the only terms  involving an explicit
appearance of $A_3$). Demanding that the action be gauge invariant
up to boundary terms only allows these two Chern-Simons terms, and
their coefficients are fixed by a combination of supersymmetry and
1-loop computations in the dimensionally reduced IIA theory.
Alternatively, the anomaly inflow computation we will now describe
can be viewed as another derivation of these coefficients.

 We now reduce the action to five dimensions in the presence  of the M5-brane.
 This gives various terms, including
 \eqn\ep{\eqalign{ 2\kappa_5^2 I&= \int\! d^5x \sqrt{g}  (R+{1 \over 4}
  G_{IJ}F^I_{\mu\nu} F^{J\mu\nu})
 +{i\over 6} \int \! C_{IJK} A^I
 \wedge F^J \wedge F^K \cr &\quad +{i \kappa_5^2 \over 192\pi^2}
  {c_{2I}p_0^I }\int\! A \wedge \Tr R \wedge R ~.}}
Here $A^I$ are 1-form potentials, obtained by expanding  $A_3 =
A^I \wedge J_I$ where $J_I$ are a  basis of $(1,1)$ forms.   The
M5-brane defines a particular magnetic charge with respect to a
linear combination of gauge fields that we have called $A$.
$p_0^I$ defined so that $p^I = \left(-{1 \over 2\pi} \int_{S^2} F
\right)p_0^I$, where  (locally) $F=dA$.    $G_{IJ}$ is a metric on
the vectormultiplet moduli space, whose form can be found in, e.g,
\KrausGH.    It turns out to be convenient to choose units with
$\kappa_5^2  =2\pi^2$, and so we do this from now on.  This
simplifies the relation  between integrally  conserved charges and
flux integrals.

Both Chern-Simons terms in \ep\ contribute to the current inflow,
and hence to the central charges \en.  By counting powers of gauge
fields we can see that the $AFF$ term yields the terms in the
central charges cubic in the $p^I$, while the $ARR$ term yields
the linear terms.

 It turns out that the cubic
term is more difficult to obtain. The idea is that one needs to
carefully define the action \ep\ in the presence of the string
source, which naively acts as a delta function source.   After
smoothing out the source and defining the AFF term appropriately,
one indeed reproduces the cubic terms in $c$ and $\ct$.  We direct
the reader to \HMM\ for the analysis.   We will shortly carry out
a corresponding analysis in the near horizon geometry of the
string, and reproduce this result in a simpler way. Right now we
just emphasize that only the $AFF$ Chern-Simons term is needed for
the result.

Turning to the $ARR$ term in the second line of \ep, we now show
how to compute the linear terms in the central charges.     We
first need to rewrite the action.  Since the string is
magnetically charged, $A$ is not globally defined.   The correct
version of the Chern-Simons term corresponds to integrating by
parts,
\eqn\eq{  I =  -  {i c_{2I}p_0^I \over 384 \pi^2 }\int\! F \wedge
\Tr(\Gamma d\Gamma
+{2\over 3} \Gamma^3)~. }
 Now perform a coordinate transformation $\delta \Gamma = dv+[\Gamma,v]$,
 \eqn\er{ \delta I =-  {ic_{2I}p_0^I \over 384 \pi^2 }\int\! F \wedge
 \Tr ( dv \wedge d\Gamma)~.}
In the present context we are thinking about the string as a
localized source, so $dF=0$ except at the string, where it is a
delta function in the transverse space.  Hence if we integrate
 \er\ by parts, as well as integrate over the transverse space,  we obtain
 \eqn\es{\delta I =- {ic_{2I}p^I \over 192 \pi} \int \!\Tr(vd\Gamma)~.}
$v$ and $\Gamma$ are $5\times 5$ matrices, since they originated
as the connection in five dimensions.   We can write them in block
diagonal form, corresponding to the conections on the tangent and
normal bundles with respect to the string worldvolume.   Taking
$v$ to act on the tangent space, the variation \es\ corresponds to
a gravitational anomaly, and by comparing with \ea-\eaa\ we read
off
\eqn\et{ c-\ct = \half c_{2I}p^I~.}
Now taking $v$ to act in the transverse space we obtain the
``normal bundle anomaly".  From the worldvolume point of view this
is the same as the $SU(2$) R-symmetry anomaly.  Relabelling in
$SU(2)$ language: $\Gamma^{ab} \rightarrow \At^{ab} =
\epsilon^{abc} \At^c$ (and similarly relabelling $v$ as
$\tilde{\Lambda}$) the variation becomes
 \eqn\eu{ \delta I = {ic_{2I}p^I \over 48\pi  } \int \Tr(\tilde{\Lambda} d\At)~.}
Comparing with \em\ we read off $\kt= c_{2I}p^I/12$, or $\ct =
\half c_{2I}p^I$. We remind the reader that this is just the
contribution linear in $p^I$. Combining this result with \et\ we
correctly reproduce the linear terms in \en.

The success of the anomaly inflow computation in reproducing the
microscopic central charges can be thought of as a consistency
check.   Any mismatch would imply that M-theory is  quantum
mechanically inconsistent in the presence of M5-branes. From a
practical standpoint, if we accept that anomalies should cancel,
the inflow method is a very efficient means of extracting the
central charges, since we only need to know the Chern-Simons terms
in the effective action, and these are highly constrained.

We now shift gears and turn to the analysis in the near horizon region.
The asymptotically flat solution of the five dimensional theory  \ep\ is
\eqn\ev{\eqalign{ds^2&=({1 \over 6}C_{IJK}H^I H^J
H^K)^{-1/3}(-dt^2+dx_4^2)+({1 \over 6}C_{IJK}H^I H^J
H^K)^{2/3}(dr^2+r^2 d\Omega_2^2) \cr A^I &= \half p^I (1+\cos
\theta)d\phi
\cr H^I& =\overline{X}^I+ {p^I \over 2r}~.}}
The vectormultiplet moduli also take nontrivial values that we
have not written out.  See, e.g., \KrausVZ.

To examine the near horizon geometry we write
\eqn\jba{r={ {1 \over 6}C_{IJK}p^I p^J p^K \over 2 z^2}~.}
For $z\rightarrow \infty$ we then find the following AdS$_3 \times
S^2 $ geometry
\eqn\jbb{\eqalign{ds^2 & =  \ell^2 {-dt^2 + dx_4^2 +dz^2 \over
z^2} + {1 \over 4} \ell^2 d\Omega_2^2
}}
with
\eqn\jbc{ \ell  =\left({1 \over 6}C_{IJK}p^I p^J
p^K\right)^{1/3}~.}
The Brown-Henneaux computation of the central charge applied to
this case gives (recalling $G_5 = \kappa_5^2/8\pi = \pi/4$)
\eqn\jbd{ c= {3\ell \over 2G_3} =   {3\pi \ell^3 \over 2G_5} =
C_{IJK}p^I p^J p^K~.}
This result, along with the form of the solution, can alternatively be derived by the method of c-extremization as described earlier.

As expected, \jbd\ yields the leading large charge contribution to
the central charge. We now turn to the AdS  computation  of the
exact central charges using anomalies. As we have discussed, this
reduces to determining the exact coefficients of the gauge and
gravitational Chern-Simons terms in the three dimensional
effective action.  That is, given
\eqn\ew{I_{CS}= {i\kt \over 4\pi }\int \Tr(\At d\At +{2 \over 3} \At^3) -i \beta \int \Tr(\Gamma
d\Gamma+{2 \over 3}\Gamma^3)}
we can read off
\eqn\ex{ \ct = 6\kt,\quad c-\ct = 96 \pi \beta~.}
The three dimensional Chern-Simons terms  descend from those in
eleven dimensions, \eo, or equivalently in five dimensions, \ep.
To read off the desired terms we can consider the following
metric deformation of AdS$_3 \times S^2$
\eqn\ey{ds^2 = ds_{AdS}^2 + {1 \over 4} \ell^2  (dy^i -
\At^{ij}y^j)(dy^i - \At^{ik}y^k)~,}
with $\sum_{i=1}^3 (y^i)^2 =1$.   This identifies  the 1-forms
$\At^{ij}$ as the $SO(3) \cong SU(2)$ gauge fields appearing in
\ew.   We also need to give the 2-form potential supporting the
solution.    From \ev\ the undeformed solution has $F^I = dA^I =
-\half p^I \epsilon_{S^2}$, where $\epsilon_{S^2}$ is the volume
form on the unit two-sphere.    We want a generalization
consistent with $SO(3)$ gauge invariance.   Since the metric is
invariant under
\eqn\ez{ \eqalign{ y^i &\rightarrow y^i +\Lambda^{ij}y^j \cr
\At^{ij}& \rightarrow \At^{ij} + d\Lambda^{ij} +[\Lambda,A]^{ij}~,}}
where $\Lambda^{ij} = -\Lambda^{ji}$ depends only on the AdS
coordinates, we also demand this of $F^I$.    $F^I$ must also be
closed and have a fixed integral over the $S^2$ fibre, since this
integral gives the 5-brane charge.   The unique solution to this
problem is \refs{\HMM,\HansenWU}
\eqn\fa{ F^I = -\half p^I (4\pi e_2)}
with
\eqn\fb{\eqalign{ e_2 &= {1 \over 8\pi} \epsilon_{ijk}(
Dy^i Dy^j- \tilde{F}^{ij}  ) y^k \cr Dy^i & = dy^i -\At^{ij}y^j \cr \tilde{F}^{ij} & = d\At^{ij}-
\At^{ik} \At^{kj}~.}}
$e_2$ is known as the ``global angular 2-form".    The $AFF$ term
in \ep\ will now yield $\At$ dependent terms.  To work these out,
a very useful formula is \Bott
\eqn\fc{ \int e_0^{(1)} \wedge e_2 \wedge e_2 =-\half  \left({1
\over 2\pi}\right)^2 \int \Tr (\At d\At +{2 \over 3} \At^3)~,}
where the integral on the left(right) is over  five(three)
dimensions.  $e_0^{(1)} $ is defined by writing $e_2 = de_0^{(1)}
$, which can always be done locally since $e_2$ is closed.    The
$AFF$ term then yields (recall $\kappa_5^2 = 2\pi^2$)
\eqn\fd{ I_{CS} = {i \over 24\pi^2} \int C_{IJK} A^I \wedge  F^J
\wedge F^K = {i \over 24\pi}C_{IJK} p^I p^J p^K \int \Tr(\At d\At
+{2 \over 3} \At^3) ~.}
 This yields the coefficient of
the Chern-Simons terms cubic in $p^I$.  Indeed,  comparing with
\ew\ and using \ex\ we correctly read off the cubic terms in the
central charges \en.

The linear terms come from the Chern-Simons term in the second
line of \ep. We can follow the same steps as led to \eq.   The
difference is that in the near horizon geometry there is no
explicit string source, but rather a smooth geometry, and so
$dF=0$ everywhere, without delta  function singularities.    After
performing the $S^2$ integration, \eq\ splits into two terms
corresponding to Chern-Simons terms for the $SO(3) \cong SU(2)$
connection, and the AdS$_3$ Christoffel connection,
\eqn\fe{I_{CS} ={ic_{2I}p^I \over 48 \pi}  \int \Tr(\At d\At +{2
\over 3}\At^3)  -{ic_{2I}p^I \over 192\pi} \int \Tr(\Gamma d\Gamma
+{2 \over 3}\Gamma^3)~. }
Note that the relative factor of $4$ between  these two terms is
purely due to our use of $SU(2)$ conventions for $\At$. From \fe\
we correctly read off the linear terms in the central charges.

We again want to stress that this computation  yields the {\it
exact} central charges, and that we did not need to know the full
action to carry it out.  Knowledge of the Chern-Simons terms
suffices, since they give us the anomalies, and supersymmetry
connects these to the central charges.  The fact that we found
exact agreement with the microscopic central charges was explained
already in terms of the anomaly inflow mechanism.    The match is
necessary to preserve diffeomorphism invariance of M-theory in the
presence of M5-branes.  Even though the agreement was guaranteed
to occur, it is still satisfying to see it working in explicit
detail.

Now that we have verified the exact matching of the microscopic
and gravitational central charges, we know that the entropy of an
uncharged black hole is given by \cs\ and that it matches with the
CFT entropy.   This matching includes subleading corrections to
the Bekenstein-Hawking area law, as encoded in the corrections to
the central charge. It further applies to non-BPS and non-extremal
black holes.   Historically, the result for the corrected entropy
of the BPS black holes  was first obtained in \CardosoFP\ by
explicitly constructing the black hole solutions in supergravity
supplemented by certain $R^2$ terms. Surprisingly, this gives the
exact result even though $R^4$ and higher type terms are not
incorporated. However, the method of \CardosoFP\ is not successful
in capturing the corrected entropy of non-BPS and non-extremal
black holes \SahooRP.

We should note that our results  so far are only valid to leading
order in $L_0$ and $\Lt_0$, and furthermore does not allow for the
inclusion of charge.  In the remainder of these lectures we will
show to how generalize in these directions.

\subsec{Small black holes and heterotic strings}

We have seen from anomalies that the bulk AdS$_3$ theory exactly
reproduces the microscopic central charges \en.  Since this result
is exact, it can be used even in cases where the bulk geometry is
highly curved and the two-derivative approximation to the action
is no longer valid.   It is especially interesting to consider
examples where the microscopic theory is as simple as possible, so
that we have good control over the microscopic entropy counting.
Such ``small black holes" have been the subject of much recent
discussion, e.g. \refs{\curvcorr,\DDMP,\KrausVZ,\Sen}

A good example is to consider $M_6 = K3\times T^2$ and to wrap the
M5-brane on $K3$.  In this case only a single magnetic charge
$p^I$ is nonzero, and hence $C_{IJK} p^I p^J p^K =0$.    This
implies that in the two derivative approximation, where
$c={3\ell\over 2G}$, the size of the AdS$_3$ geometry shrinks to
zero.   However, from \en\ we see that including higher
derivatives yields $c=24p$ and $\ct =12p$, where we used  $c_2(K3)
=24$. Strictly speaking, our supergravity analysis tells us that
{\it if} there is a finite size AdS$_3$ geometry, then its central
charges are as stated.  To actually demonstrate the existence of
the geometry require more detailed consideration of the explicit
supergravity equations of motion, including higher derivatives.
The state-of-the-art at the moment is to include just the
supersymmetric completion of certain $R^2$ terms, and to show that
a stabilized geometry indeed results \CardosoFP.   While working
out the precise solution is an important challenge, we would like
to emphasize that getting the central charges right is not too
dependent on the details, since symmetries and anomalies are
enough to determine them.

The connection with the heterotic string is obtained by using
heterotic/IIA duality.  This duality interchanges the M5-brane
(NS5-brane in the IIA language) with an elementary heterotic
string.   The magnetic  charge $p$ becomes the winding number of
the heterotic string around an $S^1$. The $24$ leftmoving
transverse bosonic oscillators of a heterotic string yield $c=12$;
and the $8$ rightmoving transverse bosonic and fermionic
oscillators yield $\ct =12$.   Taking into account the winding
number, we see precise agreement with the supergravity side. From
our discussion  so far, this means that we will find agreement in
the entropies from the Cardy formula \cs.  Note that this
agreement pertains even for non-supersymmetric and nonextremal
states (both left and right movers excited).

\newsec{Partition functions and  elliptic genera}

So far we have discussed black hole entropy at the level of the
Cardy formula.  We now try to go further in establishing the
AdS/CFT relation
\eqn\ff{Z_{AdS} = Z_{CFT}~.}
In this section we discuss the definitions and properties of the
CFT partition functions that we will subsequently aim to reproduce
from AdS.

In full generality, we can imagine defining $Z_{CFT}$ by tracing
over the CFT Hilbert space weighted by $e^{-\beta H}$ and an
arbitrary string of operators.  In principle such an object has a
dual AdS definition, but in practice it will be intractable to
actually compute. Rather than including  all possible operators,
it is more tractable to just focus on conserved charges, since
these are more easily identifiable on the gravity side.  If we
define the CFT on a circle, the two most obvious conserved charges
are energy and momentum, related to the Virasoro charges
as\foot{$P$ is the same as what we earlier called $J$, the AdS
angular momentum.}
\eqn\fg{  H = L_0 - {c\over 24} +\Lt_0 - {\ct \over 24}~,  \quad P
= L_0 - \Lt_0~.}
The most basic partition function is thus
\eqn\fh{ Z = \Tr \left[ e^{-\beta H +i\mu P}\right]  = e^{ {i\mu
(c-\ct) \over 24} } \Tr \left[ e^{2\pi i \tau ( L_0 - {c\over 24}
) - 2\pi i \taub ( \Lt_0 - {\ct \over 24})}\right]~,}
with $\tau = (\mu+i\beta)/2\pi $.   If fermions  are present we
also need to specify their periodicity around the circle.

Now suppose that our CFT also has conserved currents, $J^I$ and $\Jt^I$.
Although we use the same index $I$ for both, the number of left moving
(holomorphic) currents $J^I$ is independent of the number of right moving
(anti-holomorphic) currents $\Jt^I$.
We can generalize
our partition function by adding chemical potentials for the corresponding
conserved charges $q^I$ and $\qt^I$,
\eqn\fiz{ Z = \Tr \left[ e^{2\pi i \tau ( L_0 - {c\over 24} ) -
2\pi i \taub ( \Lt_0 -  {\ct \over 24})}e^{2\pi i z_I q^I}e^{-2\pi
i \zt_I \qt^I} \right]~.}

The path integral version of the partition function \fiz\ is,
\eqn\fia{ Z_{PI} = \int\! [{\cal D} \Phi] ~e^{-I -{i \over 2\pi}
\int (A^\mu J_\mu+ \At^\mu \Jt_\mu)}~,}
where the CFT is defined on the torus.     The external gauge
fields appearing in \fia\ are related to the chemical potentials
in  \fiz,
\eqn\fj{ z_I = -i \tau_2 A_{I \wb}~,\quad \zt_I = i \tau_2 \At_{Iw}~,}
where $\tau = \tau_1 + i\tau_2$.    Further, the the path
integral and canonical versions are related as
\eqn\fk{Z = e^{-{\pi \over \tau}(z^2 + \zt^2)}Z_{PI}~,}
with $z^2 = k^{IJ}z_I z_J$,   ($k^{IJ}$ is defined in \dq),  and
similarly for $\zt^2$.

To derive \fj-\fk, it is most instructive to consider a simple example of a
free scalar field.  This example will also allow us to discuss the modular
behavior of our partition functions.

\subsec{ Free scalar field example}

Consider a free compact boson of radius $2\pi R$. We use the
conventions of \PolchinskiRQ\ and set $\alpha'=1$. We define the
partition function
\eqn\ga{ Z(\tau,z,\zt) = (q\qb)^{-1/ 24} \Tr\left[ q^{L_0}
\qb^{\Lt_0} e^{2\pi i z p_L} e^{2\pi i \zt p_R} \right]~,}
with
\eqn\gb{\eqalign{ L_0 &= {p_L^2 \over 4} +L_0^{osc}~, \quad  \Lt_0
= {p_R^2 \over 4} +\Lt_0^{osc} \cr p_L &= {n \over R}+wR~,\quad
p_R = {n\over R}-wR~.}}
The partition function obeys the modular transformation rule
\eqn\gh{ Z({a\tau + b \over c\tau +d},{z\over c\tau +d},{\zt\over c\taub +d})
=e^{ { 2\pi icz^2 \over c \tau+d}}e^{ -{2\pi i c\zt^2 \over c\taub+d} }
Z(\tau,z,\zt)~,}
as is readily verified by direct computation.

To explain the origin of the exponential prefactors in \gh\ we pass to a path
integral formulation.   We consider
\eqn\gj{ Z_{PI}(\tau,A) = \int\! {\cal D}X e^{-I} }
with
\eqn\gk{ I = {1 \over 2\pi}\int_{T^2}d^2\sigma \sqrt{g}\left[ {1
\over 2}g^{ij}\p_i X \p_j X  -A^i \p_i X \right] }
and $A^i =$ constant.  To relate potentials appearing in \ga\ and
\gk, we use the standard expression for the charges
\eqn\gka{ p_L = 2 \oint {dw \over 2\pi i}i\p_w X~,\quad p_R = -2
\oint {dw \over 2\pi i}i\p_{\wb} X~,}
and then equate the charge dependent phases in the two versions.
This yields
\eqn\gn{ z= -i\tau_2 A_{\wb}~,\quad \zt=i\tau_2 \At_w~.}
We denoted the holomorphic part of the gauge field $\At_w$ because,
elsewhere in these notes,   this component arises from an independent bulk
1-form $\At$.

In the path integral formulation a modular transformation is a
coordinate transformation combined with a Weyl transformation, and
so it is manifest that
\eqn\gha{ Z_{PI}({a\tau + b \over c\tau +d},{z\over c\tau
+d},{\zt\over c\taub +d}) = Z_{PI}(\tau,z,\zt)~,}
where the transformation of $z$ and $\zt$ just expresses the
coordinate transformation.

What then is the relation between $Z_{PI}$ and $Z$?  To find this
we just carry out the usual steps that relate Hamiltonian and path
integral expressions (e.g.  $\int \! {\cal D}X e^{-I} = \Tr
e^{-\beta H}$.) The only point to be aware of is that the
Hamiltonian corresponding to the action \gk\ is not the factor
appearing in the exponential of \ga, but differs from this by a
contribution quadratic in the potentials, as is verified by
carrying out the standard Legendre transformation.  In particular,
we find
\eqn\ghb{ Z_{PI}(\tau,z,\zt) = e^{{\pi (z+\zt)^2 \over
\tau_2}}Z(\tau,z,\zt)~.}
Combining \gha\ and \ghb\ we see that the modular transformation law
of $Z$ must be such to precisely offset that of  $e^{{\pi (z+\zt)^2
\over \tau_2}}$.  This is what \gh\ does.

To summarize, we have shown how to convert between the canonical
and path integral versions of the partition function.  The latter makes
the modular behavior manifest.
Furthermore, the
analysis we performed is essentially completely general, in that
given an arbitrary CFT we can always realize the $U(1)$ current
algebra in terms of free bosons.

\subsec{Elliptic genus}

The partition function \fiz\ receives contributions from all
states of the theory. This makes it intractable to calculate
explicitly, except in favorable cases (such as weak coupling
limits).    In a theory with enough supersymmetry we can define a
more controlled object -- the ``elliptic genus" --  which only
receives contributions from BPS states.   The elliptic genus is a
topological invariant, as we'll review in a moment, which allows
it to be computed far more readily than the generic partition
function.  Useful references include
\refs{\KawaiJK,\farey,\MooreFG}.

For definiteness, we now focus on a CFT with $(0,4)$ susy.     The elliptic
genus is defined as
\eqn\fl{\eqalign{ \chi(\tau,z_I)&= \Tr_{R} \left[ e^{2\pi i \tau
(L_0 -c/24)} e^{-2\pi i \taub (\Lt_0-\tilde{c}/24)} e^{2\pi i z_I q^I}(-1)^{\tilde{F}}\right]
~.}}
The trace is over the Ramond sector, and $\tilde{F}$ is the
fermion number, defined as $\tilde{F} = 2\Jt^3_0$,  where
$\Jt^3_0$ is the R-charge.   The insertion of $(-1)^{\tilde{F}}$
imposes a bose-fermi cancellation among all states {\it except}
those obeying $\Lt_0-\tilde{c}/24=0$ (the Ramond ground states).
 The arguments here are the same
as in the study of the Witten index in 4D supersymmetric field theories.
Since only states   with
$\Lt_0-\tilde{c}/24=0$ contribute, the elliptic genus
does not depend explicitly on $\taub$.    On the other hand, all
leftmoving states can contribute.   The elliptic genus is invariant under
smooth deformations of the CFT.  This follow from the quantization of the
charges and of $L_0 - \Lt_0$, together with the fact that only rightmoving
ground states contribute.    We can therefore compute the elliptic genus
in the free limit of the CFT, and then extrapolate it to strong coupling and
 compare with a supergravity computation.

We now state the main general properties of the elliptic
genus.

\vskip.2cm \noindent {\bf Modular transformation}

\eqn\fm{ \chi({a\tau + b \over c\tau +d},{z_I
\over c\tau +d}) =e^{2\pi i  {c z^2 \over c\tau +d} } \chi(\tau,z_I)~.}
The same argument applies here as in \ghb.

\vskip.2cm \noindent {\bf Spectral flow}

The modes of the stress tensor and currents obey the algebra
\eqn\fn{\eqalign{ [L_m,L_n]& = (m-n)L_{m+n} +{c\over
12}(m^3-m)\delta_{m+n,0}~, \cr [L_m,J^I_{n}]&=-n J^J_{m+n}~, \cr
[J^I_{m},J^J_{n}]&=\half m k^{IJ} \delta_{m+n,0}~. }}
This is invariant under the spectral flow automorphism \dra.

The spectral flow automorphism implies the relation
\eqn\fp{\chi(\tau,z_I+\ell_I \tau+m_I)= e^{-{2\pi i}  ( \ell^2\tau
+2 \ell \cdot z)}\chi(\tau,z_I)~,
  }
where $m_I$ obeys $m_I q^I \in \IZ$, and  we defined $\ell^2 =
k^{IJ}\ell_I \ell_J$, $\ell \cdot z = k^{IJ} \ell_I z_J$.  It also
implies that if we expand the elliptic genus as
\eqn\acb{\chi(\tau,z_I) = \sum_{n, r^I} c(n,r^I)
e^{2\pi i n \tau + 2\pi i z_I r^I}~, }
then the expansion coefficients are a function of a
single spectral flow invariant combination:
\eqn\ada{ c(n,r^I) = c(n-{r^2 \over 4}) ~.}
Here we defined $r^2 = k_{IJ} r^I r^J$, where $k_{IJ}$ denotes the inverse of
$k^{IJ}$.

\vskip.2cm \noindent {\bf Factorization of dependence on
potentials}

We can explicitly write the dependence of the elliptic genus on
the potentials $z_I$.  The intuition behind this is that we
can always separate the CFT into the currents  plus everything
else, and the current part can
be realized in terms of free bosons.  We have:
\eqn\adz{\chi(\tau,z_I) = \sum_{\mu^I}
h_{\mu}(\tau) \Theta_{\mu,k}(\tau,z_I)~,}
with
\eqn\aez{\eqalign{  \Theta_{\mu,k}(\tau,z_I)&=   \sum_{\eta_I} e^{
{i\pi \tau \over 2}(\mu+2k\eta )^2}   e^{2\pi i
z_I(\mu^I+2k^{IJ}\eta_J)}~.}}
We are using the shorthand notation
\eqn\aeza{  (\mu+2k\eta )^2 \equiv k_{IJ}(\mu^I  +
2k^{IK}\eta_K)(\mu^J + 2 k^{JL}\eta_L)~.}

The combined sum over $\mu^I$ and $\eta_I$  includes the complete
spectrum of charges.  The sum over $\eta_I$ corresponds to shifts
of the charges by spectral flow, and so the sum on $\mu_I$ is over
a fundamental domain with respect to these shifts.  A more
intuitive understanding of \adz-\aez\ will emerge when we rederive
these results from the AdS side.

\vskip.2cm \noindent {\bf Farey tail expansion}

The main observation of \farey\  was that  upon applying the
``Farey tail transform", the elliptic genus admits an expansion
that is suggestive of a supergravity interpretation in terms of  a
sum over geometries.   We will essentially state the result here,
referring to \farey\ for the detailed derivation.  The CFT
discussion in \farey\ has recently been adapted to the $(0,4)$
context in \deBoerVG.

The properties \fm\ and \fp\ are the definitions of  a ``weak
Jacobi form" of weight $w=0$ and index $k$.  Actually, the
definition strictly applies when $k$ is a single number rather
than a matrix, but we will still use this langauge.

The Farey tail transformed elliptic genus is
\eqn\mb{ \chit(\tau,z_I)= \left({1 \over 2\pi i}\p_\tau-{1 \over 4}
{\p_z^2\over (2\pi i)^2}\right)^{3/2}\chi(\tau,z_I)~,}
where $\p_z^2 = k_{IJ} \p_{z_I} \p_{z_J}$.  $\chit$
is a weak Jacobi form of weight $3$ and index $k$, and admits the expansion
\eqn\mc{\chit(\tau,z_I)=e^{-{\pi z^2 \over \tau_2}}
\sum_{\Gamma_\infty\setminus\Gamma}{1\over (c\tau+d)^3
}\hat{\chi}\left( {a\tau+b\over c\tau+d},{z_I\over ct +d}\right)~,
}
with
\eqn\md{\eqalign{ \hat{\chi}(\tau,z_I) &=e^{{\pi z^2 \over \tau_2}}
\hat{\sum_{\mu,\tilde{\mu},m,\tilde{m}}}
\tilde{c}(m,\mu^I)e^{2\pi i (m-{1 \over
4}\mu^2)\tau}\Theta_{\mu,k}(\tau,z_I)~,}}
and $\Theta_{\mu,k}(\tau,z_I)$ was defined in  \aez.    The hatted
summation appearing in \md\ is over states with $m-{1 \over
4}\mu^2 <0$. From the gravitational point of view these will be
states below the black hole threshold and the sum over
$\Gamma_\infty\setminus\Gamma$ then adds the black holes back in.
In mathematical terminology \md\ defines $\hat{\chi}$ as the
``polar part" of the elliptic genus.
   The coefficients $\tilde{c}(m,\mu^I)$ in
\md\ are related  to those in \acb\ by
\eqn\mda{ \tilde{c}(m,\mu^I) = (m-{\mu^2 \over 4})^{3/2}
c(m-{\mu^2 \over 4})~,}
as follows from \mb\ and from using \ada.  The main point  is that
the transformed elliptic genus $\tilde{\chi}$ can be reconstructed
in terms of its polar part $\hat{\chi}$.

\newsec{Computation of partition functions in gravity: warmup examples}

We now turn to the gravitational computation of partition
functions, particularly the elliptic genus.  One goal will be to
see how the general properties described in the previous section
are realized in terms of the sum over geometries.  For example, we
need to see how a sum over black hole geometries, with the precise
weighting factors specified by \mc-\md,  arises in the AdS
description.

Before considering the general problem of summing over geometries, it
will be helpful to get oriented by considering some  examples.
Again, for definiteness we will focus on the $(0,4)$ case, although
the generalization to the $(4,4)$ case is very straightforward.

\subsec{NS vacuum}

The NS vacuum is invariant under $SL(2,\IR)\times  SL(2,\IR)$.  In
other words, it is invariant under the full group of AdS$_3$
isometries, which means that it is precisely global AdS$_3$,
\eqn\agc{ ds^2 = (1 +r^2/\ell^2)\ell^2dt^2  +{dr^2 \over
1+r^2/\ell^2} +r^2 d\phi^2~.}
The contractibility of the $\phi$ circle forces the fermions  to
be anti-periodic in $\phi$. Invariance under the isometry group
means that this geometry has
\eqn\agd{ L_0 = \Lt_0 =0~.}

\subsec{Spectral flow to the R sector}

On the gravity side a rightmoving spectral flow \dra\ is implemented
by a constant shift in the gauge potentials \ds, but now in terms of
the rightmoving tilded version. To get to the Ramond sector we want
to flip the periodicity of the supercurrent.  This carries charge
$\qt^0 =1$, and so we should take $\etat_0 = \half$. Therefore, a
Ramond ground states consists of the metric \agc\ with
\eqn\age{ \At_{0\wb} = 1~,}
with  fermions  periodic in $\phi$. The gauge field contribution
\do\ increases the Virasoro charge from \agd\ to
\eqn\agf{\Lt_0 = {\kt\over 4} = {\ct\over 24}~.}
Since the charge \dpp\  is
\eqn\agfa{ \qt^0 = \kt ={\ct\over 6}~,}
this is the maximally charged R vacuum state.\foot{This bound on the
charge can be seen from the supersymmetry algebra.}   To get the
maximally negatively charged R vacuum we flip the sign in \age. In
the $(4,4)$ case the leftmoving side is treated analogously.

\subsec{Conical defects}

A more general  class of R vacua are the conical defect geometries
\refs{\BalasubramanianRT,\MaldacenaDR}. For these we take
\eqn\agg{ \eqalign{ ds^2 &= ({1 \over N^2}+{r^2 \over \ell^2})dt^2
+{dr^2 \over ({1 \over N^2}+{r^2 \over \ell^2})} +r^2 d\phi^2 ~,
\cr \At_{0\wb} &={1 \over N}~,}}
with $N\in \IZ$.   The angular coordinate $\phi$  has the standard
$2\pi$ periodicity, and fermions are taken to be periodic in
$\phi$.

To read off the Virasoro charges we just note that by rescaling
coordinates all these geometries are locally equivalent to the
$N=1$ case discussed in the previous example.   In the $N=1$  case
the right moving stress tensor vanishes, and it will clearly
continue to vanish after rescaling coordinates. Thus \agf\ still
applies and so $\Lt_0 = {\kt \over 4}$ as before.    The R-charge
is read off from \do-\dpp\  as
\eqn\agh{\qt^0 = {\kt \over N}~.}
Upper and lower bounds on $N$ are given by the quantization of
R-charge, so  $|N|\leq \kt$.

These conical defect geometries are singular at the  origin unless
the holonomy is $\pm 1$, which corresponds to $N=\pm 1$.  In the
context of the D1-D5 system, the singular geometries  are known to
be  physical in that the singularity corresponds to the presence
of $N$ coincident  Kaluza-Klein monopoles.    Another way of
viewing this is that these singular geometries are special limits
of the much larger class of smooth RR vacua geometries that have
been heavily studied in recent years \refs{\LuninIZ,\MathurZP}.

We also note that any of the R-vacua in \agg\ can be spectral flowed
to the NS sector to give chiral primary geometries.

\subsec{Black holes}

We now consider black hole geometries, and give a simple derivation
of the entropy of charged black holes that incorporates higher
derivative corrections.   This will provide the generalization of
\cs.   We again use the method of relating the black hole to thermal
AdS by a modular transformation.   We will be considering a general,
rotating, non-extremal, charged black hole.  All left and right moving
charges will be turned on.

The starting point is global AdS$_3$, as in \agc. The complex
boundary coordinate is $w= \phi +i t/\ell$, and we identify
$w\cong w+ 2\pi \cong w+2\pi \tau$. To add charge we also want to
turn on flat potentials  for the gauge fields.   Now, the $\phi$
circle is contractible in the bulk, so to avoid a singularity at
the origin we need to set to zero the $\phi$ component of all
potentials.  We therefore allow nonzero $A_{Iw} = - A_{I\wb}$,
and $\At_{I\wb} = - \At_{Iw}$.

What is the action associated with this solution? From the
discussion in section 3 we know the exact expressions for the
stress tensor and currents
\eqn\ahc{\eqalign{ T_{ww}&=-{k \over 8\pi} +{1 \over 8\pi}
A_w^2+{1 \over 8\pi} \At_w^2~,  \cr T_{\wb\wb}&= -{\kt \over 8\pi}
+{1 \over 8\pi}A_{\wb}^2+{1 \over 8\pi}\At_{\wb}^2~, \cr J^I_w &=
{i\over 2} k^{IJ} A_{Jw}~, \cr \Jt^I_{\wb}& = {i\over 2} \kt^{IJ}
\At_{J\wb}~.}}
To obtain the exact action from these formulae  we need to
integrate the equation
\eqn\aha{ \delta I = \int_{\p AdS}\! d^2x \sqrt{g^{(0)}}\left( {1
\over 2}T^{\alpha\beta} \delta g^{(0)}_{\alpha\beta}+{i\over 2\pi}
J^{I\alpha} \delta A_{I\alpha}+{i\over 2\pi} \Jt^{I\alpha} \delta
\At_{I\alpha} \right)~.}
As we did to derive \ch, we first need to switch to the $z$
coordinates \cf\ that have fixed periodicities.    Doing this,
then switching back to the $w$ coordinates, we find \eqn\ahcb{
\delta I = (2\pi)^2 i \left[ -T_{ww} \delta\tau +  T_{{\bar
w}{\bar w}}\delta {\bar\tau} + {\tau_2\over\pi} J^I_w \delta
A_{I\wb} + {\tau_2\over\pi} \Jt^I_{\wb} \delta \At_{Iw}
\right]_{\rm const}~. }
The {\it const} subscript  indicates that we keep just the zero
mode part. Inserting \ahc\ into this equation we can now integrate
and find our desired action as
\eqn\ahcc{ I= {i \pi k \over 2}\tau -{i\pi \kt \over 2} \taub
+\pi \tau_2 (A_{\wb}^2 + \At_w^2)~.}

A simpler derivation of this result is to just compute \ahcc\ by
directly evaluating the action on the solution. The gauge field
contribution just comes from the boundary terms in \dn. The reason
we proceeded in terms of \aha\ was to emphasize that the result
\ahc\ is exact for an arbitrary higher derivative action, and also
because we will generalize this computation later.

The result \ahcc\ is the action for the AdS$_3$ ground state with
a flat connection turned on. Next, we perform the modular
transformation $\tau \rightarrow -1/ \tau$ in order to reinterpret
the solution as a Euclidean  black hole. This is implemented by
\eqn\ahd{ w \rightarrow -w/\tau, \quad A_{I\wb}  \rightarrow
-{\bar\tau} A_{I\wb}~,\quad \At_{Iw} \rightarrow - \tau
\At_{Iw}~.}
The action is of course invariant since we are just rewriting it in
new variables. Using $\taub/\tau = 1- 2i \tau_2/\tau$ we can present
the result as
\eqn\ahe{\eqalign{ I&= -{i\pi k \over 2 \tau}+{i\pi \kt \over
2\taub} -{2\pi i\tau_2^2 A_{\wb}^2 \over \tau} + {2\pi i\tau_2^2
\At_{\wb}^2 \over \taub}+\pi \tau_2 (A_{\wb}^2 +\At_w^2) \cr &
=-{i\pi k \over 2 \tau}+{i\pi \kt \over 2\taub} + {2\pi i z^2 \over
\tau} - {2\pi i \zt^2 \over \taub}-{\pi\over \tau_2}   (z^2+\zt^2)
~.}} This is the Euclidean action of a black hole with modular
parameter $\tau$ and potentials specified by $z_I$ and $\zt_I$.

Our result \ahe\ is the leading saddle point contribution to the
path integral. As we noted  in \fk\ the  canonical form of the
partition function, defined as a trace, is related to the path
integral as
\eqn\ahf{\eqalign{ Z &=e^{-{ \pi \over \tau_2} (z^2 +\zt^2)}  Z_{PI}
= e^{-{ \pi \over \tau_2} (z^2 +\zt^2)}\sum e^{-I}  ~.}}
The exponential prefactor cancels the last term in \ahe\ so that
\eqn\ahg{ \ln Z =  {i\pi k \over 2 \tau}-{i\pi \kt \over 2\taub} -
{2\pi i z^2 \over \tau} + {2\pi i \zt^2 \over \taub}~,}
on the saddle point. We define the entropy $s$ by writing the partition
function as
\eqn\agh{ Z = e^{s}  e^{2\pi i \tau
(L_0 -c/24)} e^{-2\pi i \taub (\Lt_0-\tilde{c}/24)} e^{2\pi i z_I q^I} e^{-2\pi i \zt_I q^I}~,}
where we assume that $Z$ is dominated by a single  a single charge
configuration with, e.g.,  $q^I = {1 \over 2\pi i } {\p \over \p
z_I  } \ln Z$.

Putting everything together we read off the black hole entropy as
\eqn\agi{ s= 2\pi \sqrt{{c \over 6} (L_0 - {c \over 24} -{1 \over 4}q^2)  }
+ 2\pi \sqrt{{\ct \over 6} (\Lt_0 - {\ct \over 24} -{1 \over 4}\qt^2)  }~. }
The expression \agi\ gives the entropy for a general nonextremal,
rotating, charged, black hole in AdS$_3$, including the  effect of
higher derivative corrections as incorporated in the central
charges. Since we used the saddle point approximation the formula
is only valid to leading order in $L_0 - {c \over 24} -{1 \over
4}q^2$ (and the rightmoving analogue); including the subleading
contribution is the topic of the next section.  It is striking
that we have control over higher derivative corrections to the
entropy even for nonsupersymmetric black holes.\foot{A related
observation is that the attractor mechanism, which plays an
important role in establishing a near horizon AdS$_3$ geometry,
can also operate for non-supersymmetric black holes \GoldsteinHQ.}
As in our discussion of the uncharged case, the relation with
anomalies implies that \agi\ is in precise agreement with the
microscopic entropy counting coming from brane constructions.

\newsec{Computation of partition functions in supergravity}

Let's now look at the supergravity computation of the elliptic
genus.  We'll consider both the canonical and path integral
approaches, which are useful for making manifest the behavior
under spectral flow and modular transformation, respectively. In
keeping with the Farey tail philosophy \farey, we first explicitly
compute the contribution to the elliptic genus from states below
the black hole threshold. With this in hand, we then note that
black holes are readily included since they are just coordinate
transformations of solutions below the threshold.   In this way we
reproduce the construction \mc.

\subsec{Canonical approach}

In the canonical approach we need to enumerate the allowed set of
bulk solutions and their charge assignments.    For the elliptic
genus we consider states of the form (anything, R-ground state),
which have $\Lt_0 = {\kt\over 4}$.   There are three classes of such
states: smooth solutions in the effective three dimensional theory;
states coming from Kaluza-Klein reduction of the higher dimensional
supergravity theory; and non-supergravity string/brane states.
Some members of the first class were discussed above, and we
will make a few comments on  the other types of states later.

Just as was done on the CFT side \adz, it is useful to factorize
the dependence on the potentials. In the gravitational context it is
manifest that the stress tensor consists of a metric part plus  a
gauge field part.   Suppose we are given a state carrying leftmoving
charges
\eqn\agia{ (L_0-{c\over 24}, q^I)=(m,\mu^I)~. } We can apply
spectral flow to generate the family of states with charges
\eqn\agi{\eqalign{L_0-{c\over 24} &=   m +\eta_I
q^I+k^{IJ}\eta_I\eta_J = m-{1 \over 4}\mu^2 +{1 \over 4}
(\mu+2k\eta)^2 \cr   q^I&= \mu^I +2 k^{IJ} \eta_J~,}}
where we are using the same shorthand notation as in \aeza.   This
class of states will then contribute to the elliptic genus as
\eqn\agj{ \chi(\tau,z_I) = (-1)^{\tilde{F}} e^{2\pi i \tau (m-{1 \over
4}\mu^2)} \Theta_{\mu,k}(\tau,z_I)~,}
in terms of the $\Theta$-function \aez. Each such spectral flow
orbit has a certain degeneracy from the number distinct states with
these charges.   We call this degeneracy $c(m-{1 \over 4}\mu^2)$,
where the functional dependence is fixed by the spectral flow
invariance, and we also include $(-1)^{\tilde{F}}$ in the definition.    We
can now write down the ``polar" part of the elliptic genus, that is,
the contribution below the black hole threshold: $m-{1 \over 4}\mu^2
<0$.   We then have
\eqn\agk{ \chi^\prime(\tau,z_I) = \sum_{m,\mu}\nolimits^{\prime}
c(m-{1 \over 4}\mu^2) \Theta_{\mu,k}(\tau,z_I) e^{2\pi i (m-{1 \over
4}\mu^2)\tau} ~.}

In the canonical  approach it is easy to write down the polar part
of the elliptic genus in terms of the degeneracies $c(m-{1 \over
4}\mu^2)$.   But the full elliptic genus also has a contribution
from black holes, and these are not easily incorporated since black
holes do not correspond to individual states of the theory.  To
incorporate black holes we need to turn to a Euclidean path
integral, as we do now.

\subsec{Path integral approach}

In the path integral approach we sum over bulk solutions with fixed
boundary conditions
\eqn\agl{\chi_{PI}(\tau,z_I) = \sum e^{-I}~.}
The action appearing in \agl\ is the full string/M-theory
effective action reduced to AdS$_3$, though we fortunately do not
require its explicit form to compute the elliptic genus.  In
particular, in \agl\ we only sum over stationary points of $I$
since the fluctuations have already been incorporated through
higher derivative corrections to the action.

The boundary conditions on the metric are that the boundary geometry
is a torus of modular parameter $\tau$.   $z_I$ fix the boundary
conditions for the gauge potentials.  As derived in \gn,
the relation is, in conformal gauge,
\eqn\agm{A_{I\wb} = {iz_I \over \tau_2}~.}
$A_{Iw}$ is not fixed as a boundary condition.   Since the potential
$\zt_I$ is set to zero in the elliptic genus, we also have the
boundary condition
\eqn\agn{ \At_{Iw}=0~.}

Now we turn to the allowed values of $A_{Iw}$ and $\At_{I\wb}$.
The allowed boundary values of $A_{Iw}$ are determined from the
holonomies around the contractible cycle of the AdS$_3$ geometry.
Recall that when we write $w=\sigma_1+i\sigma_2$ we are taking
$\sigma_1$ to be the $2\pi$ periodic spatial angular coordinate. The
corresponding cycle on the boundary torus is contractible in the
bulk, and so any nonzero holonomy must match onto an appropriate
source in order to be physical. The holonomy of a charge $q^I$
particle is
\eqn\agq{ e^{\half i q^I \int\! d\sigma_1 A_{I\sigma_1} } =e^{ \half
i q^I \int\!  d\sigma_1 (A_{Iw} +A_{I\wb})}~.}
Choosing a gauge with constant $A_{Iw}$, we write the allowed values
as
\eqn\agr{ A_{Iw} = k_{IJ} \mu^I +2\eta_I -{iz_I \over \tau_2}~,\quad
q^I\eta_I \in \IZ~,}
where we have written the charge of the source as $\mu^I$.

In the same way we can determine the allowed values of $\At_{I\wb}$.
In this case we know that only geometries with $\Lt_0- {\ct \over
24}=0$ contribute to the elliptic genus, and so we do not include
the spectral flowed geometries as we did above.  Instead, we just
have
\eqn\agu{ \At_{I\wb} = \kt_{IJ} \mut^I~.}
%

%
%

Given the gauge fields, we know the exact stress tensor  and also
the exact currents.   We can therefore find the action by
integrating
\eqn\ago{\eqalign{ \delta I &= \int_{\p AdS}\! d^2x
\sqrt{g^{(0)}}\left( {1 \over 2}T^{\alpha\beta} \delta
g^{(0)}_{\alpha\beta}+{i\over 2\pi} J^{I\alpha} \delta A_{I\alpha}
\right) \cr &= (2\pi)^2 i \left[ -T_{ww} \delta\tau + T_{{\bar
w}{\bar w}}\delta {\bar\tau} + {\tau_2\over\pi} J^I_w \delta
A_{I\wb} + {\tau_2\over\pi} \Jt^I_{\wb} \delta \At_{I w}
\right]_{\rm const} ~,}}
as in section 4.4. The result is
\eqn\agt{\eqalign{ I&=  -2\pi i \tau(L_0^{grav}-{c\over 24})+2\pi
i \taub (\Lt_0^{grav}-{\ct\over 24})\cr &\quad-{i\pi  \over
2}\left[ \tau A^2_{w}  +\taub A^2_{\wb}  + 2\taub A_{w} A_{\wb}
\right] +{i\pi  \over 2} \left[ \tau \At^2_{w}  +\taub \At^2_{\wb}
+ 2\tau \At_{w} \At_{\wb} \right]~. }}
In verifying that \agt\ satisfies \ago\ one has to take care to
consider only variations consistent with the equations of motion and
the assumed boundary conditions. We maintain fixed holonomies by
taking $\delta A_{Iw} = - \delta A_{I\wb}$ and $\delta \At_{Iw} = -
\delta\At_{I\wb}$. Also, the variation of the complex structure must
be taken with the gauge field fixed in the $z$-coordinates
introduced in \cf.

The result \agt\ for the action agrees with \ahc\ when the
geometry is in the ground state where $A_{Iw}\ = - A_{I\wb}$ and
$\At_{Iw}\ = - \At_{I\wb}$, but it is valid also more generally in
the presence of charged sources. In fact, it is equivalent to the
canonical result discussed in section 7.1. To see this we consider
again the charge assignments \agia. Writing $L_0 = L_0^{grav}  +
L_0^{gauge} = L_0^{grav} + {1 \over 4}\mu^2$ (and analogously for
$\Lt_0$) we insert into \agt\ and find
\eqn\agua{ I = -2\pi i \tau(m-{1 \over 4} \mu^2) -{i\pi  \tau\over
2} (\mu+2k \eta)^2-2\pi i  z_I (\mu^I+2k^{IJ}\eta_J)-{\pi  z^2 \over
\tau_2}~. }
Summing over the geometries below the black hole threshold we find
\eqn\agv{\eqalign{ \chi'_{PI}(\tau,z_I) &=
\sum\nolimits^{\prime}_{m,\mu} c(m-{1 \over 4}\mu^2) e^{-S} \cr & =
e^{{\pi  z^2 \over \tau_2}}\sum\nolimits^{\prime}_{m,\mu} c(m-{1
\over 4} \mu^2) \Theta_{\mu,k}(\tau,z_I) e^{2\pi i (m-{1 \over 4}
\mu^2)\tau} \cr & = e^{{\pi  z^2 \over \tau_2}}  \chi'(\tau,z_I)~,}}
where $\chi^\prime$ is the canonical result \agk. As in  \fk, the
overall exponential factor is precisely the one we expect.

\subsec{Including black holes}

Black holes are readily included in the path integral approach since
they are just rewritten versions of solutions below the black hole
threshold.  Taking a solution below the black threshold and
performing the coordinate transformation $w\rightarrow {aw + b \over
cw +d}$ generates  a black hole.   Using the manifest invariance of
the action under such  coordinate transformations, the contribution
of such a black is then
\eqn\agw{ \chi_{PI}(\tau,z_I) = \chi'_{PI}\left( {a\tau +b \over
c\tau +d}, {z_I \over c\tau+d}\right)~. }
On the other hand, from the relation \agv\ between $\chi'_{PI}$ and
$\chi'$ we have
\eqn\agx{ \chi'_{PI}\left( {a\tau +b \over c\tau +d}, {z_I \over
c\tau+d}\right) = e^{-2\pi i {c z^2 \over c\tau +d}}e^{{\pi z^2
\over \tau_2}}\chi' \left( {a\tau +b \over c\tau +d}, {z_I \over
c\tau+d}\right)~.}
Thus the black hole contribution to $\chi$ is
\eqn\agy{ \chi(\tau,z_I)= e^{-{\pi z^2 \over \tau_2}}
\chi_{PI}(\tau,z_I) = e^{-2\pi i {c z^2 \over c\tau +d}}\chi' \left(
{a\tau +b \over c\tau +d}, {z_I \over c\tau+d}\right)~.}

The next step is to sum over all inequivalent black holes to get the
complete elliptic genus.  This means summing over the subgroup of
$\Gamma=SL(2,\IZ)$ corresponding to inequivalent black holes or,
more precisely, distinct ways of labelling the contractible cycle in
terms of time and space coordinates.  As explained in \farey\ the
inequivalent cycles are parameterized by $\Gamma_\infty \setminus
\Gamma$; so it seems natural to write
\eqn\agz{ \chi(\tau,z_I)=\sum_{\Gamma_\infty \setminus \Gamma}
e^{-2\pi i {c z^2 \over c\tau +d}}\chi' \left( {a\tau +b \over c\tau
+d}, {z_I \over c\tau+d}\right)~.}
However, as emphasized in \farey, this cannot be correct since the
sum is not convergent. Instead we should compute not the elliptic
genus but instead its Farey transform, introduced in \mb. This
amounts to  first replacing $\chi'$ by
\eqn\aia{ \hat{\chi}'(\tau,z_I) =\sum\nolimits^{\prime}_{m,\mu}
\ct(m-{1 \over 4} \mu^2) \Theta_{\mu,k}(\tau,z_I) e^{2\pi i (m-{1
\over 4} \mu^2)\tau} }
with $\ct$ defined as in \mda.   We interpret this as the polar part
of a weak Jacobi form of weight $3$ and index $k$.   Instead of
\agz\ we therefore write
\eqn\aib{  \hat{\chi}(\tau,z_I)=\sum_{\Gamma_\infty \setminus
\Gamma} (c\tau+d)^{-3} e^{-2\pi i {c z^2 \over c\tau +d}}\hat{\chi}'
\left( {a\tau +b \over c\tau +d}, {z_I \over c\tau+d}\right)~.}

\subsec{High temperature behavior} The high temperature
($\tau_2\rightarrow 0$) behavior of \aib\ is governed by the free
energy of a BPS black hole.  The leading exponential behavior  can
be read off from the term
\eqn\ka{ \left(\matrix{a&b\cr c&d}\right) =\left(\matrix{0&-1\cr
1&0}\right)~,\quad m=0~,\quad \eta_I =0, \quad \mu^I = k
\delta^{I0}~,}
which gives
\eqn\kb{ \hat{\chi}(\tau,z_I) \approx  e^{-{2\pi i z^2 \over \tau}
+{2\pi i k z_0 \over \tau }  } ~.}
We can compare with \ahg\ by performing the spectral flow
$z_0\rightarrow z_0+\half$. This yields
\eqn\kba{ \ln \hat{\chi}(\tau,z_I)  \approx  {i\pi k \over 2\tau}
-{2\pi i z^2 \over \tau}~.}
Noting that this agrees with the holomorphic part of \ahg, we find
that the corresponding entropy is is indeed that of a BPS black
hole,
\eqn\kbb{ s= 2\pi \sqrt{{c \over 6} (L_0 - {c \over 24} -{1 \over
4}q^2)  } ~.}
This is just the leading part of the entropy, and is insensitive to
the distinction between the elliptic genus and its Farey-tail
transformed version.

\subsec{Summary}

It is now helpful to summarize what has been achieved so far. In
our CFT discussion we noted that the CFT elliptic genus\foot{Here
when we say elliptic genus we really mean its Farey tail
transform.} is completely determined by the spectrum of BPS states
below the black hole threshold, and by the algebra of CFT
currents.  By evaluating the  Euclidean path integral, we then
showed that the AdS elliptic genus has precisely the same
structure.   Thus we have boiled the question of exact agreement
of the elliptic genera to the comparison of current algebras and
BPS states below the black hole threshold.  To complete the
computation these need to be worked out.  Some aspects of this
problem on the AdS side are the subject of the next section.

\newsec{Computation of BPS spectra}

There are in general two types of BPS states to consider:
supergravity states from the Kaluza-Klein fluctuation spectrum of
the higher dimensional theory reduced to AdS$_3$; and branes
wrapping cycles  of the internal compactification manifold. We do
not intend to give a full description of either here, and restrict
ourselves to sketching some aspects.

\subsec{Supergravity states}

First consider the supergravity fluctuations.  The starting point
is either eleven dimensional supergravity on AdS$_3 \times S^2
\times M_6$, or IIB supergravity on AdS$_3 \times S^3 \times M_4$.
For definiteness we focus on the former; the approach in the two
cases is very similar. After reduction on $M_6$ one has a five
dimensional supergravity theory with some number of
vectormultiplets $n_V$; hypermultiplets $n_H$; and gravitino
multiplets $n_S$;\foot{Gravitino multiplets are present for
$M_6=T^6$ or $K3\times T^2$ to capture the extra supersymmetry.}
in addition to the gravity multiplet.     The multiplicities of
each multiplet are determined by the Hodge numbers of $M_6$, and
are sumamrized in Table 1.
\bigskip
\vbox{
$$\vbox{\offinterlineskip
\hrule height 1.1pt
\halign{&\vrule width 1.1pt#
&\strut\quad#\hfil\quad&
\vrule width 1.1pt#
&\strut\quad#\hfil\quad&
\vrule width 1.1pt#
&\strut\quad#\hfil\quad&
\vrule width 1.1pt#
&\strut\quad#\hfil\quad&
\vrule width 1.1pt#\cr
height3pt
&\omit&
&\omit&
&\omit&
&\omit&
\cr
&\hfil $M_6$ &
&\hfil $n_S$&
&\hfil $n_V $&
&\hfil $n_H$ &
\cr
height3pt
&\omit&
&\omit&
&\omit&
&\omit&
\cr
\noalign{\hrule height 1.1pt}
height3pt
&\omit&
&\omit&
&\omit&
&\omit&
\cr
&\hfil $CY_3$ &
&\hfil $0$&
&\hfil $h^{1,1}-1 $&
&\hfil $2h^{1,2}+2$&
\cr
height3pt
&\omit&
&\omit&
&\omit&
&\omit&
\cr
\noalign{\hrule}
height3pt
&\omit&
&\omit&
&\omit&
&\omit&
\cr
&\hfil $K3\times T^2$ &
&\hfil $2$&
&\hfil $22$&
&\hfil $42$&
\cr
\noalign{\hrule}
height3pt
&\omit&
&\omit&
&\omit&
&\omit&
\cr
&\hfil $T^6$ &
&\hfil $6$&
&\hfil $14$&
&\hfil $14$&
\cr
&\omit&
&\omit&
&\omit&
&\omit&
\cr
}\hrule height 1.1pt
}
$$
}
\centerline{\sl Table 1: 5-dimensional supergravity spectra.}

\def\sugrastates{Table 2}
\bigskip

The next step is to expand in harmonics on the $S^2$, to get an
AdS$_3$ spectrum of fields.  The modes appearing in the expansion
of each field yields a representation of the symmetry group, which
includes the subgroup $SL(2,\IR)_L \times SL(2,\IR)_R \times
SU(2)_R$ corresponding to the AdS$_3$ isometries and the
R-symmetry.   For the computation of the elliptic genus we just
need the spectrum of chiral primaries, which are those modes
obeying $\tilde{h} = \half \qt^0$, where $\Lt_0=\tilde{h}$ and
$\Jt^3_0 = \half \qt^0$ are the $SL(2,\IR)_L$ and $SU(2)_R$
weights.  The value of $L_0 =h$ is unrestricted by the chiral
primary condition, and indeed we can generate a whole tower by
application of $L_{-1}$ to a lowest $h$ state.   The details of
the  computation of this spectrum can be found in
\KutasovZH\foot{The earlier references \refs{\finn,\deBoerIP} give
incorrect ranges of ${\bar h}$ that differ slightly from these.};
we summarize the result in \sugrastates. Since the chiral
primaries form multiplets under $SL(2,\IR)_L$ symmetry, in
\sugrastates\ we list the spectrum of single particle chiral
primaries that are also primary under the leftmoving $SL(2,\IR)$;
i.e. are annihilated by $L_{1}$.
\bigskip
\vbox{
$$\vbox{\offinterlineskip
\hrule height 1.1pt \halign{&\vrule width 1.1pt#
&\strut\quad#\hfil\quad& \vrule width 1.1pt#
&\strut\quad#\hfil\quad& \vrule width 1.1pt#
&\strut\quad#\hfil\quad& \vrule width 1.1pt#\cr height3pt &\omit&
&\omit& &\omit& \cr &\hfil $s=h-\tilde{h}$& &\hfil degeneracy& &\hfil
range of $\tilde{h}=\half \qt^0$& \cr height3pt &\omit& &\omit& &\omit&
\cr \noalign{\hrule height 1.1pt} height3pt &\omit& &\omit&
&\omit& \cr &\hfil $1/2$& &\hfil $n_H$& &\hfil $1/2,3/2,\ldots$&
\cr height3pt &\omit& &\omit& &\omit& \cr \noalign{\hrule}
height3pt &\omit& &\omit& &\omit& \cr &\hfil $0$& &\hfil $n_V$&
&\hfil $1,2,\ldots$& \cr \noalign{\hrule} height3pt &\omit&
&\omit& &\omit& \cr &\hfil $1$& &\hfil $n_V$& &\hfil $1,2,\ldots$&
\cr \noalign{\hrule} height3pt &\omit& &\omit& &\omit& \cr &\hfil
$-1/2$& &\hfil $n_S$& &\hfil $3/2,5/2,\ldots$& \cr
\noalign{\hrule} height3pt &\omit& &\omit& &\omit& \cr &\hfil
$1/2$& &\hfil $n_S$& &\hfil $3/2,5/2,\ldots$& \cr \noalign{\hrule}
height3pt &\omit& &\omit& &\omit& \cr &\hfil $3/2$& &\hfil $n_S$&
&\hfil $1/2,3/2,\ldots$& \cr \noalign{\hrule} height3pt &\omit&
&\omit& &\omit& \cr &\hfil $-1$& &\hfil $1$& &\hfil $2,3,\ldots$&
\cr \noalign{\hrule} height3pt &\omit& &\omit& &\omit& \cr &\hfil
$0$& &\hfil $1$& &\hfil $2,3,\ldots$& \cr \noalign{\hrule}
height3pt &\omit& &\omit& &\omit& \cr &\hfil $1$& &\hfil $1$&
&\hfil $1,2,\ldots$& \cr \noalign{\hrule} height3pt &\omit&
&\omit& &\omit& \cr &\hfil $2$& &\hfil $1$& &\hfil $1,2,\ldots$&
\cr height3pt &\omit& &\omit& &\omit& \cr }\hrule height 1.1pt }
$$
} \centerline{\sl Table 2: Spectrum of (non-singleton) chiral
primaries for $AdS_3\times S^2\times M_6$
.}
\bigskip
The tower of $\tilde{h}$ values correspond to the tower of spherical harmonics
on $S^2$.   This spectrum does not include the so-called singletons; we'll come
back to this point momentarily.

Given this spectrum it is straightforward to work out the elliptic genus as
\eqn\zza{\chi^{sugra} = \Tr_{chir.~prim.}  \left[ (-1)^{\qt^0}
q^{L_0} \right] }
where $q= e^{2\pi i \tau}$.  The sum over states  includes
multiparticle contributions. The result is
\eqn\zzb{ \chi^{sugra}(\tau) = M(q)^{-{\rm
Euler}}\prod_{n=1}^\infty (1-q^n)^{n_v+3-2n_s}(1-q^{n+1})~,  }
where the McMahon function is defined as
\eqn\zzc{ M(q) = \prod_{n=1}^\infty (1-q^n)^n~,}
and ``Euler" denotes the Euler number of $M_6$.

Now we incorporate the singletons. Singleton modes are pure gauge
configurations that are nonetheless physical in the presence of
the AdS$_3$ boundary.   To see why, consider the case of a $U(1)$
gauge field with Chern-Simons term.  The configuration $A_w = \p_w
\Lambda(w)$ is formally pure gauge, but from \dm\ it carries the
nonzero stress tensor $T_{ww} = {k \over 8\pi} (\p_w \Lambda)^2$,
and hence is physical. This is possible because the true gauge
transformations must vanish at the boundary and it is only those
that leave the stress tensor invariant. The singleton states are
described in the CFT as $J_{-1}|0\rangle$, where $J$ is the
current  corresponding to $A$.   We also have the $SL(2,\IR)$
descendants of these states.

A similar story holds for singletons associated with
diffeomorphisms that are nonvanishing at the boundary.     These
correspond to  the states $L_{-2}|0\rangle$ and $SL(2,\IR)$
descendants thereof.   The explicit form of the diffeomorphisms is
given in \brownhen.

We can now work out the contribution of the singletons to the elliptic genus
of the $(0,4)$ theory.    If there
are $n_L$ leftmoving currents then the contribution of singletons is
\eqn\mj{\chi_{NS}^{sing} = \prod_{n=1}^\infty {1 \over
(1-q^n)^{n_L}}{1 \over ( 1-q^{n+1}) }~. }

We need to know the number of leftmoving currents, which involves
knowing the form of the AdS$_3$ Chern-Simons terms.  These can be
worked out from reduction of the eleven dimensional theory, and
gives
\eqn\mk{ n_L= \left\{ \matrix{5 & T^6 \cr 21 & K3\times T^2 \cr
n_V & CY_3 } \right.~.  }
See \KrausNB\ for the derivation.

We find the full result by multiplying \zzb\ and \mj:
\eqn\mka{ \chi_{NS} =\chi_{NS}^{sugra} \chi_{NS}^{sing}=   \left\{
\matrix{1 & T^6 \cr &\cr  1 & K3\times T^2 \cr & \cr M(q)^{-{\rm
Euler}}\prod_{n=1}^\infty (1-q^n)^{3} & CY_3 } \right.  }
We find that in the $T^6$ and $K3\times T^2$ cases the singletons
precisely cancel the dynamical contribution \zzb. For the  CY$_3$
the dependence on $n_V$ cancelled.  Note that these conclusion are
a result of cancellations between propagating states from
\sugrastates\ and the singletons.

\subsec{Contribution from wrapped branes}

The final ingredient in the computation of the elliptic genus is
the contribution from wrapped branes.    In the $(0,4)$ theory
corresponding to M-theory on AdS$_3 \times S^2 \times M_6$ these
are M2-branes wrapped on 2-cycles of $M_6$.    In \GaiottoNS\ it
was shown that this computation is equivalent to the
Gopakumar-Vafa derivation \GopakumarII\  of the topological string
partition function from   M-theory, and this leads to the
connection between the black hole elliptic genus and the
topological string.  The main novelty is that both M2-branes and
anti-M2-branes turn out to preserve the same supersymmetry when
situated at opposite poles of the $S^2$.   The complete
contribution then takes the form of an absolute square, which in
turn leads to the OSV relation between $Z_{BH}$ and $|Z_{top}|^2$.
There is much more that can be said here, but we refer the reader
to \refs{\GaiottoNS,\deBoerVG} for more details.

To bring the story to its logical conclusion, one should now try
to make the explicit comparison with the $(0,4)$ CFT, analogous to
what was done in \deBoerUS.   This requires an explicit result for
the CFT elliptic genus, which is not available so far.   We again
refer the reader to the references for what is presently known.

\newsec{Conclusion}

We hope to have given the reader an understanding of how to
compute  the entropy  of an AdS$_3$ black hole, and compare with
CFT. One main lesson is that the success of most of the black hole
/ CFT entropy comparisons in the literature can be traced back to
the matching of symmetries and anomalies.  This gives a better
understanding of {\it why} the entropies agree, even at the
subleading level, and for certain non-supersymmetric black holes.
We have also sketched the route by which one can hope to make {\it
exact} comparisons between black hole and CFT partition functions,
although much work remains to be done to bring this program to
completion.

We conclude by mentioning a few open issues.   In section 4.3 we
discussed how the entropies  of fundamental heterotic strings can
be deduced from a gravitational computation.   The reader might be
puzzled as to why we didn't also consider the seemingly simpler
example of type II fundamental strings.  In fact, the type II
case, rather than being simpler, is enigmatic.  From the point of
view of higher derivative terms in the spacetime action, the
difference between the two cases is that $R^2$ corrections are
absent in the type II case.  But it is such $R^2$ terms that
resolve the naked singularity of the heterotic string, replacing
it by a finite size horizon.   One also sees a crucial difference
in our anomaly based approach.  In the type II case spacetime
rotations couple non-chirally to the string worldsheet, hence
there is no anomaly inflow mechanism by which one can deduce the
central charges.  We are therefore unable to compute the entropy
on the gravitational side.    It is an important open problem as
to whether higher derivative terms (e.g. $R^4$ terms) resolve the
naked singularity of the type II string, and whether the
microscopic  entropy can be reproduced.

Finally, one of the main motivations for undertaking an extensive
study of black entropy in string theory is to shed light on the
resolution of the information paradox.   The success of the
AdS/CFT correspondence is usually interpreted to mean that there
is no information loss, since the boundary CFT has manifestly
unitary evolution, and so one can in principle track the explicit
time evolution of any given microstate.  A truly satisfying
resolution of the information paradox will involve providing an
analogous description in the bulk.   In the context of the
computations described here, we would like to be able to compute
the AdS partition functions via an explicit sum over bulk states.
The tools for such a computation are currently being developed in
the context of deriving bulk states dual to CFT microstates (for
reviews see \refs{\MathurZP,\iosif}). It will be very illuminating
to see how the same AdS partition function can be computed either
by summing over black hole geometries, or by enumerating
individual bulk states.

\bigskip
\noindent {\bf Acknowledgments:} \medskip \noindent I thank
Stefano Bellucci for the invitation to lecture at the Frascati
winter school, and the participants for stimulating discussions.
Thanks also  to Finn Larsen and James Hansen for collaboration in
the work described here.   This work was supported in part by NSF
grant PHY-00-99590.

 \listrefs

\end